%% file: benchmarksg.tex
\documentclass[letterpaper,11pt]{article}
\usepackage{epsfig}
\input{basic.tex}
\def\Eisen{G_{2}\(t,\bar{t}\)}
\def\modelnum{seven}

\begin{document}

\begin{center}
            \hfill    Fermilab-PUB-02/211-T \\
            \hfill    MCTP-02-46 \\

\end{center}

\bigskip

\begin{center}
{\large \bf THEORY-MOTIVATED BENCHMARK MODELS AND SUPERPARTNERS AT
THE TEVATRON}
\end{center}

\begin{center}
G.~L.~Kane$^{1}$, J.~Lykken$^{2}$, Stephen~Mrenna$^{2}$,
Brent~D.~Nelson$^{1}$, Lian-Tao~Wang$^{1}$ and Ting~T.~Wang$^{1}$
\end{center}

\begin{center}
$^{1}${\it Michigan Center for Theoretical Physics, Randall
Lab.,\\

University of Michigan, Ann Arbor, MI 48109}
\end{center}

\begin{center}
$^{2}${\it Theoretical Physics Department,\\

Fermi National Accelerator Laboratory, Batavia, IL 60510}
\end{center}

\begin{quotation}
Recently published benchmark models have contained rather heavy
superpartners. To test the robustness of this result, several
benchmark models have been constructed based on theoretically
well-motivated approaches, particularly string-based ones. These
include variations on anomaly and gauge-mediated models, as well
as gravity mediation. The resulting spectra often have light
gauginos that are produced in significant quantities at the
Tevatron collider, or will be at a 500 GeV linear collider. The
signatures also provide interesting challenges for the LHC. In
addition, these models are capable of accounting for electroweak
symmetry breaking with less severe cancellations among soft
supersymmetry breaking parameters than previous benchmark models.
\end{quotation}

\section*{Introduction}

Benchmark models can be of great value in helping plan and execute
experimental analyses. They allow quantitative studies of detector
design and triggers, and can be important in setting priorities
for experimental groups. They suggest what signatures can be the
most fruitful search channels for finding new physics. For
example, if benchmark models suggest rates and signatures that
imply some kinds of new physics are unlikely to be seen compared
to others, Ph.D. thesis topics and the associated experimental
efforts may move in the direction indicated by those suggestions.
Benchmark models can also provide essential guidance about what
backgrounds are important to understand and what systematic errors
need to be controlled. Consequently it is very important that the
benchmark models not misrepresent the true physics situation.
Finally, constructing benchmark models can also be valuable
theoretical exercises, helping us to gain insight into which
features of the theory imply certain phenomena and vice versa.

To be precise, we define a benchmark model as one in the framework
of softly-broken supersymmetry and based on a theoretically
motivated high scale approach. Currently, such models cannot be
specified in sufficient detail to calculate a meaningful spectrum
of interactions without making some assumptions or approximations
-- and these should be ones which make sense in the context of the
theory.

In the past two years some benchmark models for supersymmetric
spectra and signatures have been published~\cite{bench,snowmass}.
A general and perhaps surprising feature of these benchmarks is
that the resulting superpartners are rather heavy, and in
particular few or no superpartners are likely to be observed at
the Tevatron collider. The published benchmark models are
constructed using various assumptions. Such assumptions may or may
not be true, and it is important to understand whether other
approaches to benchmark models generally lead to such heavy
spectra or not. One important concern with the published models is
that they can be consistent with electroweak symmetry breaking
(EWSB) only by having large cancellations between large
contributions to $M_{Z}.$ That is a worrisome
property~\cite{KaLyNeWa02}, and leads one to question the
relevance and implications of such models.

We have studied these issues and constructed several benchmark
models that have good underlying theoretical motivation. We find
that the resulting spectra typically do have some light
superpartners that will be produced in significant quantities at
the Tevatron or a $500 \GeV$ linear collider. Further, these
models typically do describe EWSB without large cancellations, so
perhaps their implications (including the opportunity to observe
superpartners at the Tevatron) should be taken very seriously. In
those cases where it is physically reasonable we have indicated
which parameters can be varied to provided so-called ``model
lines.''

To be explicit, we propose \modelnum~sets of high--scale
supersymmetry--breaking parameters as inputs to determine the
weak--scale properties.  All of these sets have a string theory
basis, as is explained in
Sections~\ref{sec:theory1}-\ref{sec:theory3} of the paper. To
summarize, the first set of benchmarks is motivated by
nonperturbative methods of achieving dilaton stabilization leading
to a reasonable minimum of the supersymmetry--breaking potential.
They are specified by:
\begin{eqnarray}
{\rm Case\; A:} \qquad  \lbr \tan\beta,\; m_{3/2},\; a_{\rm np}
\rbr & = & \lbr 10,\; 1500 \GeV,\;  1/15.77 \rbr
\\
{\rm Case\; B:} \qquad \lbr \tan\beta,\; m_{3/2},\; a_{\rm np}
\rbr & = & \lbr 5,\; 3200 \GeV,\;  1/37.05 \rbr
\\
{\rm Case\; C:} \qquad \lbr \tan\beta,\; m_{3/2},\; a_{\rm np}
\rbr & = & \lbr 5,\; 4300 \GeV,\;  1/61.36 \rbr,
\end{eqnarray}
where $\tan\beta$ is the ratio of Higgs vacuum expectation values,
$m_{3/2}$ is the gravitino mass, and $a_{\rm np}$ is related to
the nonperturbative corrections to the dilaton potential.
A possible model line is to vary the parameter $a_{\rm np}$ with
all other parameters fixed.
The next set of benchmark points are based on string models where
the moduli fields are responsible for breaking supersymmetry. They
are specified by $\tan\beta$, $m_{3/2}$, a Green-Schwarz
coefficient $\delta_{\GS}$, and a moduli expectation value:
\begin{eqnarray}
{\rm Case\; D:} \qquad \lbr \tan\beta,\; m_{3/2},\;
\delta_{\GS},\; \lang \re \; t\rang \rbr & = & \lbr 45,\; 20
\TeV,\; -15, \; 1.10 \rbr \\ {\rm Case\; E:} \qquad \lbr
\tan\beta,\; m_{3/2},\; \delta_{\GS},\; \lang \re \; t \rang \rbr
& = & \lbr 30,\; 20 \TeV,\; -9, \; 1.23 \rbr .
\end{eqnarray}
A possible model line is to vary $\lang \re \; t \rang$ for a
given value of $\delta_{\GS}$.
The last set of benchmarks points are based on the idea of partial
gauge--mediation arising from high--scale fields, and are
specified by $\tan\beta$, $m_{3/2}$ and the SM quantum numbers of
the high--scale fields:
\begin{equation}
{\rm Case\; F:} \qquad  \lbr \tan\beta,\; m_{3/2},\; n_D, \; n_L,
\; N_1 \rbr  =  \lbr 10,\; 120 \GeV,\;  4, \; 0, \; (3/5) \rbr
\end{equation}
\begin{equation}
{\rm Case\; G:} \qquad  \lbr \tan\beta,\; m_{3/2},\;  n_D, \; n_L,
\; N_1 \rbr  =  \lbr 20,\; 130 \GeV,\; 3, \; 0, \; (3/5) \rbr.
\end{equation}
A possible model line is the variation of quantum numbers, in
particular the hypercharge $N_1$.
The corresponding values of the ``usual'' soft terms are collected
for all the benchmark points in Table~\ref{tbl:inputs} in
Section~\ref{sec:phenom}.  The reader interested mainly in the
phenomenological implications of these benchmarks may proceed
directly to that point, especially on a first reading. The
appropriate input parameters to the {\tt{PYTHIA}} event generator
are also available \cite{webpage}.

Most previous benchmarks were based on the so-called Constrained
Minimal Supersymmetric Standard Model (cMSSM) which is
characterized by universal values $m_0$ for the soft scalar
masses, a universal gaugino mass denoted $m_{1/2}$ and a universal
trilinear scalar coupling
$A_0$~\cite{KaKoRoWe94a,KaKoRoWe94b,deGrGlKa98,deHuGlKa01,ElNaOl01,ElFaGaOl01,ElHeOlWe01},
subjected to theoretical and experimental constraints. These
models are quite simple and well defined, and could once rightly
claim to represent the state of the art in effective theory
constructions motivated by string theory. But recent progress in
realistic low-energy effective models, coupled with the recently
obtained one loop expressions for soft terms in supergravity
theories derived from strings~\cite{GaNe00b}, suggests that this
universal paradigm may not accurately reflect the underlying
string theory. In addition, phenomenological constraints that may
not hold have been imposed on these benchmark models, such as
insisting that they provide the entire cold dark matter of the
universe with the needed relic density arising only from thermal
production mechanisms.

The phenomenology of cMSSM models is largely generic. Once gaugino
mass degeneracy is enforced experimental limits on chargino masses
imply a heavy gluino; proper EWSB then requires a large
cancellation between the $\mu$ parameter and this large $M_3$. The
lower bound on the Higgs boson mass adds additional constraints.
This same pattern emerges in both gauge-mediated and
anomaly-mediated models, as they are typically constructed in the
literature~\cite{DiThWe97,GhGiWe99}. In both cases the gluino soft
mass $M_3$ is larger than the wino soft mass $M_2$ by the ratio of
gauge couplings (in the case of gauge mediation) or by the ratio
of beta-function coefficients (in the case of anomaly mediation).
The LEP bound on the chargino mass again implies a heavy gluino
and proper EWSB demands once more a large value for the $\mu$
term.

We will utilize the complete one-loop expression for soft
supersymmetry breaking parameters in order to investigate three
classes of string-derived low-energy models. All of these examples
will rely on significant contributions to various soft
supersymmetry breaking terms from supergravity loop corrections,
including those that arise via the superconformal anomaly. Our
first two examples explore the implications of the two leading
methods known for stabilizing the string dilaton. Our third set of
models investigates the possibility that supersymmetry breaking is
transmitted from the hidden to the observable sector through the
agency of the standard moduli fields of string-derived
supergravity as well as vector-like multiplets of chiral
superfields charged under the gauge groups of the Standard Model.
Such exotic states are a common feature of string models and they
will necessarily give rise to a ``partial'' gauge mediation of
supersymmetry breaking. Phenomenologically, we only require that
the models are consistent with all collider data and not
significantly inconsistent with indirect constraints -- since some
constraints typically imposed in phenomenological studies (such as
thermal relic densities for LSP neutralinos) are model-dependent
and/or sensitive to input parameters, we impose these constraints
somewhat loosely. All models preserve gauge coupling unification.
We discuss details of how EWSB occurs in each case.

In Sections~\ref{sec:theory1} through~\ref{sec:theory3} we
describe in some detail the theoretical construction of the
models. While we are not arguing that any of them are
overwhelmingly compelling, we describe them in sufficient detail
so that the reader can see they are theoretically well-motivated.
In Section~\ref{sec:phenom} we present the resulting spectra.
There we will briefly summarize some phenomenological aspects of
the models, including a few observations about Tevatron signatures
and rates and a discussion on fine-tuning in these models, before
concluding. We have collected the complete one-loop expressions
for the soft terms used in this study in the Appendix, where we
indicate the set of free parameters in each case and suggest
certain model lines for further inquiry. Since the models we
construct are interesting theoretically beyond their role as
benchmark models, and for the most part have not been studied to
date, we include both theoretical descriptions of these models as
well as numerical results in the same paper. Readers who are
mainly interested in the spectra and/or the high scale input
parameters can focus on Section~\ref{sec:phenom} and
Tables~\ref{tbl:inputs} and~\ref{tbl:spectraSUM} contained
therein.

\section{K\"ahler stabilization of the dilaton}
\label{sec:theory1}

\input{theory1g.tex}

\section{Untwisted moduli-domination with multiple condensates}
\label{sec:theory2}

\input{theory2g.tex}

\section{Partial gauge mediation}
\label{sec:theory3}

\input{theory3g.tex}

\section{Collected spectra and signatures}
\label{sec:phenom}

\input{phenomg.tex}

\section*{Conclusion}

In many ways effective field theories derived from strings are at
once more constrained and also richer in their phenomenology than
the universal scenario of minimal supergravity upon which so many
previous benchmark cases are based. They are richer in that
patterns of nonuniversality, particularly in the gaugino mass
sector, are quite common; they are more constrained in that these
patterns and hierarchies are not completely free for the
model-builder to choose but are a function of the string moduli
space. We have deliberately sought out models which imply
superpartners that are observable at the Tevatron, but we did not
have to search far: such models are common from effective field
theories derived from the weakly coupled heterotic string.

The resulting benchmark models are interesting to study, both for
theorists who want to improve our understanding of how to relate
string theory and the real world (or who wish to make progress
towards a string-derived supersymmetric Standard Model), and also
for experimentalists who wish to learn how to detect
supersymmetric signals.

\section*{Acknowledgements}

The authors would like to acknowledge helpful conversations with,
and suggestions from, Z.~Chacko, T.~Dent, L.~Everett, J.~Giedt,
J.~Wells and E.~Witten. BDN would like to thank the Lawrence
Berkeley National Laboratory for hospitality during the early
portions of this work.

\section*{Appendix}

In this Appendix we present the complete expressions for the soft
supersymmetry breaking terms at one loop in modular-invariant
supergravity theories derived from string theory. We take special
care to describe the various contributions to the gaugino masses,
as they play a special role in the text. For scalar masses and
trilinear A-terms we merely give the results. More details can be
found in~\cite{BiGaNe01}.

To obtain the soft supersymmetry-breaking Lagrangian in a
string-based model, the first step is the construction of the
four-dimensional effective supergravity theory by a dimensional
reduction of the ten-dimensional supergravity theory representing
the superstring~\cite{Wi85,FeKoPo86,CvLoOv88}. Such a procedure
yields the K\"ahler potential, superpotential, and gauge kinetic
function for the effective supergravity theory. Of particular
importance for the question of supersymmetry breaking are the
types of string moduli present in the low-energy theory and their
couplings to the observable fields of the
MSSM~\cite{CvLoOv88,DiKaLo90,DiKaLo91}. Gaugino masses will depend
on auxiliary fields related to moduli appearing in the gauge
kinetic function, while scalar masses, trilinear A-terms and
bilinear B-terms will depend on auxiliary fields related to those
moduli that appear in the superpotential couplings and/or K\"ahler
potential for the MSSM fields~\cite{KaLo93,BrIbMu94}. The precise
form of these soft terms can be obtained by working out the
component Lagrangian for the observable sector by standard
techniques~\cite{CrFeGiVa83,WessBagger,BiGiGr01}.

To begin with, we take the K\"ahler potential for the moduli
fields to be given by the leading-order result
\begin{equation}
K(S,T) = k(S+\oline{S})-3\ln(T+\oline{T}). \label{modK}
\end{equation}
The tree level soft terms for the case with universal modular
weights $n_i = -1$ for all light observable sector matter fields
$Z^i$ are given by\footnote{We will not distinguish with separate
notation fields and their vacuum expectation values in these
expressions.}
\begin{eqnarray}
M_{a}^{0}&=&\frac{g_{a}^{2}}{2}F^{S} \nonumber \\ A_{ijk}^{0} &=&
-K_{s}F^{S} \nonumber \\ (m_{i}^{0})^{2}&=&\frac{M\oline{M}}{9} -
\frac{|F^{T} |^{2}}{(t + \bar{t})^{2}} . \label{treeBIM}
\end{eqnarray}

For the one-loop corrections, we begin with gaugino masses which
receive corrections from light field theory loops as well as
string loop effects. The field theory loop contribution can be
derived completely from the superconformal anomaly and is given
by~\cite{GaNeWu99,BaMoPo00}
\begin{equation} M^{1}_a|_{\rm an} =
\frac{g_{a}^{2}(\mu)}{2} \left[ \frac{2 b_a}{3} \oline{M} -
\frac{1}{8\pi^2} \left( C_a - \sum_i C^i_a \right) F^n  K_{n} -
\frac{1}{4\pi^2} \sum_i C^i_a F^n \partial_n  \ln \kappa_i
\right], \label{Man}
\end{equation}
where $C_a$, $C_a^i$ are the quadratic Casimir operators for the
gauge group ${\cal G}_a$ in the adjoint representation and in the
representation of $Z^i$. Here $b_{a}$ is given by~(\ref{ba}) in
the text and $\kappa_i$ is defined by~(\ref{kappa}).

As mentioned in the Section~\ref{sec:theory2} one expects modular
anomaly cancellation to occur through a universal Green-Schwarz
counterterm with group-independent coefficient $\delta_{\GS}$.
Such a term can be thought of as a loop-correction that
contributes to gaugino masses in the form
\begin{equation} M^{1}_a|_{\GS}
= \frac{g^2_a(\mu)}{2}  \frac{2F^T}{(t + \bar{t})}
\frac{\delta_{\GS}}{16\pi^2} . \label{MGSapx}
\end{equation}
In addition there may be string threshold corrections to the
effective gauge kinetic functions of the form
\begin{equation}
f^{1}_{a}(Z^n) = \ln \eta^2(T) \[\frac{\delta_{\GS}}{16\pi^2}  +
b_a \], \label{floop}
\end{equation}
which generate one-loop contributions to gaugino masses given by
\begin{equation} M^{1}_a|_{\rm th} =
\frac{g^{2}_{a}(\mu)}{2}\[\frac{\delta_{\GS}}{16\pi^2}+ b_a\]
 4 \zeta(t)F^{T}. \label{Mthapx}
\end{equation}
Combining the contributions from~(\ref{MGSapx}) and~(\ref{Mthapx})
with the field theory loop contribution~(\ref{Man}) gives
\begin{equation}
M_{a} = \frac{g_{a}^{2}\(\mu\)}{2} \lbr 2
 \[ \frac{\delta_{\GS}}{16\pi^{2}} + b_{a}
\]\Eisen F^{T} + \frac{2}{3}b_{a}\oline{M} +\[ 1
- 2 b_{a}' K_s \] F^{S} \rbr \label{Maloopapx}
\end{equation}
where we have defined the quantity
\begin{equation}
b_{a}' = \frac{1}{16\pi^2} \(C_a - \sum_i C_a^i \) \label{baprime}
\end{equation}
and the Eisenstein function $\Eisen$ is defined
by~(\ref{Eisenstein}) in the text.

The one-loop soft scalar masses and trilinear couplings depend on
the Pauli-Villars (PV) sector for regulating the
theory~\cite{BiGaNe01}. Here we make the simplest possible
assumption that the PV masses are constants ({\em i.e.}
independent of the moduli fields). This assumption is similar in
spirit to taking a straight cut-off represented by the
Pauli-Villars mass scale $\mu_{\PV}$. Then the soft terms in the
scalar potential are given by the $p=0$ limit of
\begin{eqnarray}
A_{ijk} &=& -\frac{K_s}{3}F^S - \frac{1}{3} \gamma_{i}\oline{M} -
p \gamma_{i} \Eisen F^{T} + \tilde{\gamma}_{i} F^{S} \lbr
\ln(\mu_{\PV}^{2}/\mu_R^2) -p\ln\[(t+\bar{t}) |\eta(t)|^4\] \rbr +
{\rm cyclic}(ijk) \nonumber \\ m_{i}^{2} &=& \lbr \frac{|M|^2}{9}
-\frac{|F^T|^{2}}{(t+\bar{t})^{2}}\rbr \[ 1 + p\gamma_i
-\(\sum_{a}\gamma_{i}^{a} -2\sum_{jk}\gamma_{i}^{jk}\) \(
\ln(\mu_{\PV}^{2}/\mu_R^2) -p\ln\[(t+\bar{t}) |\eta(t)|^4\] \) \]
\nonumber \\
 & & + (1-p)\gamma_i\frac{|M|^2}{9} + \lbr
 p\wtd{\gamma}_{i}\frac{MF^S}{6}+\hc \rbr +\lbr \wtd{\gamma}_{i}\Eisen
 \frac{\oline{F}^{T}F^S}{2} + \hc \rbr \nonumber \\
 & & +|F^{S}|^2 \[ \(\frac{3}{4}\sum_{a} \gamma_{i}^{a} g_{a}^{4}
 + K_s K_{\bar{s}} \sum_{jk} \gamma_{i}^{jk}\) \(
\ln(\mu_{\PV}^{2}/\mu_R^2) -p\ln\[(t+\bar{t}) |\eta(t)|^4\] \) \],
\label{finalsoft}
\end{eqnarray}
where we have defined the quantity $\wtd{\gamma}_{i}$ for
notational simplicity as
\begin{equation}
\wtd{\gamma}_{i} = \sum_{a} \gamma_{i}^{a} g_{a}^{2} - K_{s}
\sum_{jk} \gamma_{i}^{jk} . \label{tildegamma}
\end{equation}
The anomalous dimensions $\gamma_{i}$ are defined by
\begin{eqnarray}
\gamma^j_i = \frac{1}{32\pi^2}\[4\delta^j_i\sum_a g^2_a(T^2_a)^i_i
- e^K\sum_{kl}W_{ikl}\oline{W}^{jkl}\] . \label{gam}
\end{eqnarray}
The approximation that generational mixing can be neglected so
that only third-generation Yukawa couplings are relevant motivates
the definitions
\begin{eqnarray}
\gamma_i^j &\approx&\gamma_i\delta^j_i, \quad \gamma_i =
\sum_{jk}\gamma_i^{jk} + \sum_a\gamma_i^a, \nonumber \\ \gamma_i^a
&=& \frac{g^2_a}{8\pi^2}(T^2_a)^i_i, \quad \gamma_i^{jk} =
-\frac{e^K}{32\pi^2}
(\kappa_i\kappa_j\kappa_k)^{-1}\left|W_{ijk}\right|^2.\label{diag}
\end{eqnarray}
Thus we have
\begin{eqnarray}
(16\pi^2)\gamma_{Q_3}&=&\frac{8}{3}g_{3}^{2} +\frac{3}{2}g_{2}^{2}
+\frac{1}{30}g_{1}^{2} -\lambda_{t}^{2} -\lambda_{b}^{2} \nonumber
\\ (16\pi^2)\gamma_{U_3}&=&\frac{8}{3}g_{3}^{2}
+\frac{8}{15}g_{1}^{2} -2\lambda_{t}^{2} \nonumber
\\ (16\pi^2)\gamma_{D_3}&=&\frac{8}{3}g_{3}^{2}
+\frac{2}{15}g_{1}^{2} -2\lambda_{b}^{2} \nonumber
\\ (16\pi^2)\gamma_{L_3}&=&\frac{3}{2}g_{2}^{2}
+\frac{3}{10}g_{1}^{2} -\lambda_{\tau}^{2} \nonumber
\\ (16\pi^2)\gamma_{E_3}&=& \frac{6}{5}g_{1}^{2}
-2\lambda_{\tau}^{2}\nonumber
\\ (16\pi^2)\gamma_{H_u}&=&\frac{3}{2}g_{2}^{2}
+\frac{3}{10}g_{1}^{2} -3\lambda_{t}^{2} \nonumber
\\ (16\pi^2)\gamma_{H_d}&=&\frac{3}{2}g_{2}^{2}
+\frac{3}{10}g_{1}^{2} -3\lambda_{b}^{2} -\lambda_{\tau}^{2} ,
\label{actualgammas}
\end{eqnarray}
dropping the Yukawa coupling terms for chiral superfields of the
first and second generation. For all the models considered in this
paper we have taken $K_s = -g_{\STR}^{2}/2$ and assumed that the
regularization scale $\mu_{\PV}$ and the boundary condition scale
(here identified with $\mu_R$) coincide. This is a reasonable
approximation when the boundary condition scale is near the string
scale.

To obtain the explicit values of the soft terms that were used in
Sections~\ref{sec:theory1} and~\ref{sec:theory2} one must
substitute the appropriate expressions for the auxiliary fields
$F^S$, $F^T$ and $M$ into~(\ref{Maloopapx}) and~(\ref{finalsoft}).
For example, the model of Section~\ref{sec:theory1} is obtained by
the substitutions
\begin{eqnarray}
M&=& -3m_{3/2} \nonumber \\ F^S&=&\sqrt{3}m_{3/2}a_{\rm np}
(K_{s\bar{s}}^{\rm tree})^{-1/2} =
\frac{2\sqrt{3}}{g_{s}^{2}}a_{\rm np}m_{3/2} \nonumber \\ F^T&=&0
. \label{sec1terms}
\end{eqnarray}
A possible model line for further study is to vary the parameter
$a_{\rm np}$ over its allowed range for a particular scale
($m_{3/2}$) and value of $\tan\beta$.

The models of Section~\ref{sec:theory2} were obtained from the
same expressions as Section~\ref{sec:theory1} but with the choices
\begin{eqnarray}
M&=& -3m_{3/2} \nonumber \\ F^S&=&0 \nonumber \\ F^T&=&(t+\bar{t})
m_{3/2} . \label{sec2terms}
\end{eqnarray}
Interesting model lines for this class of models can be obtained
by continuously varying $\lang \re \; t \rang$ for various values
of the GS coefficient $\delta_{\GS}$. Such model lines have the
power to interpolate between patterns of soft supersymmetry
breaking that look similar to minimal supergravity and those that
have the features of anomaly mediation.

To obtain the models of Section~\ref{sec:theory3} one adds the
corrections given in~(\ref{gauginomass}) and~(\ref{scalarmass}) to
the soft terms of~(\ref{Maloopapx}) and~(\ref{finalsoft}) and
substitutes for the auxiliary fields $F^S$, $F^T$, $F^X$ and $M$.
Our final examples of cases~F and~G are obtained by the
substitutions
\begin{eqnarray}
M&=& -3m_{3/2} \nonumber \\
F^S&=&\frac{2\sqrt{3}}{g_{s}^{2}}a_{\rm np}m_{3/2} \nonumber \\
F^T&=&0 \nonumber \\ F^X &=& k F^S \label{sec3Bterms}
\end{eqnarray}
with $a_{\rm np} =1$ and the parameter $k$ fixed to the value
$k=1$. While both $\lang \re \; t \rang$ and $a_{\rm np}$ are free
parameters which can be varied, as are the messenger mass scale
and the phenomenological parameter $k$, a fruitful area of further
investigation is to vary the hypercharge messenger index $N_1$ for
a fixed combination of messenger indices $N_3$ and $N_2$. Like the
variable $\lang \re \; t \rang$, this parameter directly
influences the ratio $M_1/M_2$ and thus can interpolate between
minimal supergravity and anomaly-mediated spectra. For example,
the models selected in this paper for benchmark scenarios were
chosen with low values of $N_1$ to put them in the mSUGRA regime,
which is a far less challenging regime for detectors at hadronic
colliders than those of the anomaly-mediated regime.

\end{document}

%% file: basic.tex
\DeclareMathAlphabet   {\mathsc}{OT1}{cmr}{m}{sc}

\def\[{\left [}
\def\]{\right ]}
\def\({\left (}
\def\){\right )}
\newcommand{\lang}{\left\langle}
\newcommand{\rang}{\right\rangle}
\newcommand{\lbr}{\left\{}
\newcommand{\rbr}{\right\}}
\newcommand{\oline}[1]{\overline{#1}}

\newcommand{\wtd}[1]{\widetilde{#1}}

\newcommand{\GeV}      {~\mathrm{GeV}}
\newcommand{\TeV}      {~\mathrm{TeV}}

\newcommand{\EM}       {\mathsc{em}}

\newcommand{\UV}       {\mathsc{uv}}
\newcommand{\GS}       {\mathsc{gs}}
\newcommand{\PL}       {\mathsc{pl}}
\newcommand{\PV}       {\mathsc{pv}}
\newcommand{\GUT}      {\mathsc{gut}}
\newcommand{\STR}      {\mathsc{str}}

\newcommand{\lowest}{|_{\theta =\bar{\theta}=0}}

\newcommand{\hc}       {\mathrm{\; h.c. \;}}

\newcommand{\order}{\mathcal{O}}
\newcommand{\re}{{\rm Re}}

\newcommand{\gappeq}{\mathrel{\rlap {\raise.5ex\hbox{$>$}}
{\lower.5ex\hbox{$\sim$}}}}
\newcommand{\lappeq}{\mathrel{\rlap{\raise.5ex\hbox{$<$}}
{\lower.5ex\hbox{$\sim$}}}}

%% file: theory1g.tex
\subsection{Theoretical motivation}
\label{sec:motiv1}

The dilaton is a uniquely important field in string-derived
effective theories. It is the only one of the various possible
string moduli fields that always appears in the low-energy theory
in a uniform way. It represents the tree-level value of the gauge
kinetic function $f_a$ and thus its vacuum expectation value
determines the string coupling constant. In the formulation in
which the dilaton field is contained within a chiral multiplet $S$
we have
\begin{equation}
f_a^{(0)} = S; \qquad <\re \; s>\; = 1/g_{\STR}^{2} , \label{fa}
\end{equation}
where $s=S\lowest$ and $g_{\STR}$ is the universal gauge coupling
at the string scale.\footnote{We assume affine level one for
nonabelian gauge groups and 5/3 for the abelian group $U(1)_Y$ of
the Standard Model.} Though the string scale in the weakly coupled
heterotic string is typically somewhat larger than the traditional
GUT scale $\Lambda_{\GUT} \simeq 2 \times 10^{16}
\GeV$~\cite{Ka88}, we nonetheless often take the apparent
unification of gauge couplings in the MSSM as a guide and assume
$g_{\STR}^{2} \simeq 1/2$.

From~(\ref{fa}) it is clear that the low-energy phenomenology
depends crucially on finding a dynamical mechanism that ensures a
finite vacuum value for the dilaton at the observed coupling
strength. However, the superpotential for the dilaton is vanishing
at the classical level so only nonperturbative effects, of string
and/or field-theoretic origin, can create a superpotential capable
of stabilizing the dilaton~\cite{BaDi94}. There are two commonly
employed classes of solutions to this challenge~\cite{DiSh01}. The
first, sometimes referred to as ``K\"ahler stabilization,''
assumes that the tree level K\"ahler potential for the dilaton,
which is known to be of the form $K_{\rm tree}(S,\oline{S}) = -\ln
(S+\oline{S})$, is augmented by nonperturbative corrections of a
stringy origin. Then in the presence of one or more gaugino
condensates in the hidden sector the dilaton can be stabilized at
$g_{\STR}^2 = 1/2$ with a vanishing vacuum energy. This method
requires correctly choosing parameters in the postulated
nonperturbative K\"ahler potential.

The second approach, sometimes referred to as the ``racetrack''
method, assumes only the tree level form of the dilaton K\"ahler
potential but relies on at least two gaugino condensates in the
hidden sector to generate the necessary dilaton superpotential.
Generally the vacuum energy remains nonzero in such scenarios, so
some other sector must be tacitly postulated to bring about a
vanishing cosmological constant. This method requires correctly
choosing the relative sizes of the beta-function coefficients for
two different condensing gauge groups. Remarkably, concrete
manifestations of both of these two approaches -- models that have
{\em explicit} mechanisms to break supersymmetry, obtain the
appropriate dilaton vacuum value and arrange soft terms at the TeV
scale -- tend to generate nonuniversal gaugino masses and allow
for the prospect of superpartner production at the Tevatron. We
will briefly describe the first method of nonperturbative
corrections to the dilaton K\"ahler potential in this section, and
investigate the multiple condensate scenario with tree-level
K\"ahler potential in Section~\ref{sec:theory2}.

Let us begin with a brief review of the important broad features
of what Casas~\cite{Ca96} referred to as the ``generalized
dilaton-dominated'' scenario. Consider the scalar potential that
arises from any generic supergravity theory
\begin{equation}
V= K_{I\bar{J}} F^I  \oline{F}^{\bar{J}} - \frac{1}{3} M \oline{M}
\label{pot}
\end{equation}
where $F^I$ is the auxiliary field associated with the chiral
superfield $Z^I$ and $M$ is the auxiliary field of supergravity.
Note that we have suppressed the Planck mass by setting the
$M_{\PL} = 1$ here and throughout. The auxiliary fields can be
identified by their equations of motion
\begin{equation}
F^M = - e^{K/2} K^{M\oline{N}} \left(\oline{W}_{\oline{N}} +
K_{\oline{N}} \oline{W} \right), \; \; \oline{M} = -3e^{K/2}
\oline{W} \label{EQM}
\end{equation}
with $W_{\oline{N}} = \partial W / \partial \oline{Z}^{\bar{N}}$,
$K_{\oline{N}} = \partial K / \partial \oline{Z}^{\bar{N}}$ and
$K^{M \bar{N}}$ being the inverse of the K\"ahler metric $K_{M
\oline{N}}=
\partial^2 K / \partial Z^M \partial \oline{Z}^{\bar{N}}$. The
gravitino mass is given by $m_{3/2} = -\frac{1}{3}\lang
\oline{M}\rang$. The effect we wish to consider involves the
dilaton, so let us focus on the case where only the dilaton
auxiliary field $F^S$ receives a vacuum expectation value. Then
the potential~(\ref{pot}) can be written
\begin{equation}
V = K_{s\bar{s}} |F^S|^{2} -3e^{K}|W|^{2} = e^{K} K^{s\bar{s}}
|W_s + K_sW|^{2} -3e^{K}|W|^2 . \label{dilpot}
\end{equation}
We now depart from the standard case so often considered in the
literature, for which $K(S,\oline{S}) = K_{\rm tree}(S,\oline{S})
= -\ln (S+\oline{S})$ and instead allow the functions
$K_{s\bar{s}}$ and $K_s$ to be undetermined at this point.
Requiring that the potential~(\ref{dilpot}) be vanishing in the
vacuum $\lang V\rang=0$ then implies (up to an overall phase)
\begin{equation}
F^{S} = \sqrt{3} m_{3/2} (K_{s\bar{s}})^{-1/2} =
\sqrt{3}m_{3/2}a_{\rm np}(K_{s\bar{s}}^{\rm tree})^{-1/2} ,
\label{FS}
\end{equation}
where we have introduced the parameter
\begin{equation}
a_{\rm np} \equiv \(\frac{K_{s\bar{s}}^{\rm
tree}}{K_{s\bar{s}}^{\rm true}}\)^{1/2} \label{acond}
\end{equation}
designed to measure the departure of the dilaton K\"ahler
potential from its tree level value due to nonperturbative effects
of string origin. Recall that $\lang (K_{s\bar{s}}^{\rm
tree})^{1/2} \rang = \lang 1/(s+\bar{s}) \rang = g_{\STR}^{2}/2
\simeq 1/4$.

To understand the likely magnitude of the phenomenological
parameter $a_{\rm np}$ let us make the quite well-grounded
assumption that the superpotential for the dilaton is generated by
the field-theoretic nonperturbative phenomenon of gaugino
condensation and that its dilaton dependence is given by $W(S)
\propto e^{-3S/2b_{a}}$. Here $b_{a}$ is the beta-function
coefficient of a condensing gauge group ${\cal G}_{a}$ of the
hidden sector with
\begin{equation}
b_a = \frac{1}{16\pi^2} \( 3 C_a - \sum_i C_a^i \),  \label{ba}
\end{equation}
where $C_a$, $C_a^i$ are the quadratic Casimir operators for the
gauge group ${\cal G}_a$, respectively, in the adjoint
representation and in the representation of the matter fields
$Z^i$ charged under that group. Let us assume a single condensing
gauge group, which we will denote by ${\cal G}_{+}$, so that we
can write $W_s = -(3/2b_{+})W(S)$. The values of $b_+$ can be
quite a bit larger than analogous values for the Standard Model
groups, but a limiting case is that of a single $E_8$ gauge group
condensing in the hidden sector, so that ${\cal G}_{+} = {\cal
G}_{E_8}$ and $b_+ = 90/16\pi^2 = 0.57$. In what follows the
parameter $b_{+}$ will take several different values depending on
the assumed condensing gauge group. Clearly we must insist $b_+
> 0$ in order for gaugino condensation to happen at all.

Returning for a moment to the tree level case, we can now see that
requiring $\lang V \rang=0$ in~(\ref{dilpot}) would require the
following relation (understood to be taken in terms of vacuum
expectation values)
\begin{equation}
(s+\bar{s})^{2} \left| \frac{3}{2b_{+}} +\frac{1}{s+\bar{s}}
\right|^{2} = 3 \; \; \to (s+\bar{s})=\frac{2b_{+}}{3}(\sqrt{3}-1)
,
\end{equation}
and this implies $g_{\STR}^{2} \sim 1/b_{+} \sim 16\pi^{2}$. Hence
the origin of the belief that one condensate cannot stabilize the
dilaton with vanishing vacuum energy without resorting to strong
coupling. However, if we do not insist on the tree level dilaton
K\"ahler potential then the vanishing of the vacuum energy implies
\begin{equation}
(K_{s\bar{s}})^{-1}\left|K_s - \frac{3}{2b_{+}} \right|^{2} =3 \;
\; \to (K_{s\bar{s}})^{-1/2} = \sqrt{3}
\frac{\frac{2}{3}b_{+}}{1-\frac{2}{3}b_{+}K_{s}} . \label{Ktrue}
\end{equation}
So provided $K_s \sim \order(1)$ so that $K_s b_{+} \ll 1$ we can
immediately see that a K\"ahler potential which stabilizes the
dilaton while simultaneously providing zero vacuum energy will
{\em necessarily} result in a suppressed dilaton contribution to
soft supersymmetry breaking. Indeed, from~(\ref{acond})
\begin{equation}
a_{\rm np}=\sqrt{3}\frac{\frac{2}{3}\frac{g_{s}^{2}}{2}b_{+}}{1-
\frac{2}{3}K_{s}b_{+}} \ll 1 . \label{aBGW}
\end{equation}
Note that we have so far been working with only $F^{S} \neq 0$ for
the sake of simplicity. The result~(\ref{aBGW}) does not rely on
the dilaton being the only source of supersymmetry breaking ({\em
i.e.} one could always introduce more auxiliary fields $F^{X}$
with Goldstino angles in the manner of~\cite{BrIbMu94}), though
the phenomenological ramifications of~(\ref{aBGW}), to which we
will turn in the Section~\ref{sec:soft1}, will necessarily be most
pronounced when the dilaton is the dominant source of
supersymmetry breaking in the observable sector.

\subsection{A concrete realization}
\label{sec:concrete1}

Can an explicit form for the dilaton K\"ahler potential be found
that stabilizes the dilaton at values such that $g_{\STR}^{2}
\simeq 1/2$ and simultaneously providing for $\lang V \rang =0$
via the mechanism of~(\ref{Ktrue})? The question was answered
affirmatively in~\cite{BiGaWu96,BiGaWu97a} where the linear
multiplet formalism for the dilaton was employed. In this case the
dilaton field $\ell$ is the lowest component of a linear
superfield $L$  and the gauge coupling is determined by
\begin{equation}
g_{\STR}^{2} =\lang \frac{2\ell}{1+f(\ell)} \rang \label{lingauge}
\end{equation}
where $f(\ell)=f(L)\lowest$ parameterizes stringy nonperturbative
corrections to the dilaton action. This translates into a
correction to the K\"ahler potential $K(L)$ for the dilaton of the
form
\begin{equation}
K(L) = k_{\rm tree}(L) + g(L) = \ln(L) +g(L) \label{Kdil}
\end{equation}
where $g(L)$ is related to $f(L)$ by the requirement that the
Einstein term in the supergravity action have canonical
normalization, which implies:
\begin{equation}
L\frac{dg(L)}{dL} = -L\frac{df(L)}{dL} +f(L). \label{diffeq}
\end{equation}
Note that at tree level the chiral and linear multiplet
formulations are related\footnote{One should exercise extreme care
in converting from the chiral to the linear multiplet formulation,
particularly in the presence of loop corrections. For a precise
conversion of quantities such as $K_s$ and $K_{s\bar{s}}$ see
Appendix A of~\cite{BiGaNe01}} by $L=1/(S+\oline{S})$.

The form of the nonperturbative correction $f(\ell)$ used
in~\cite{BiGaWu96,BiGaWu97a} was that originally motivated by
Shenker~\cite{Sh90}
\begin{equation}
f(\ell) = \sum_{n} A_n (\sqrt{\ell})^{-n} e^{-B/\sqrt{\ell}}
\label{nonpertsum}
\end{equation}
and subsequently studied by other authors~\cite{Po94,Si97}. It is
an important feature of~(\ref{nonpertsum}) that these string
instanton effects scale like $e^{-1/g}$ (when we use $\ell \sim
g^2$) and are thus stronger than analogous nonperturbative effects
in field theory which have the form $e^{-1/g^2}$. Thus they can be
of significance even in cases where the effective four-dimensional
gauge coupling at the string scale is weak~\cite{BaDi94}.

To achieve a minimum with the desired properties it is sufficient
to truncate the expression~(\ref{nonpertsum}) after two terms and
write
\begin{equation}
f(\ell) = (A_0 + A_{1}/\sqrt{\ell}) e^{-B/\sqrt{\ell}}.
\label{nonpert}
\end{equation}
It was shown in~\cite{BiGaWu97a,GaNe00a} that such a function can
indeed stabilize the dilaton at weak coupling and vanishing vacuum
energy with $\order(1)$ parameters.\footnote{This was confirmed by
subsequent authors. See for example~\cite{BadeCo98}
where~(\ref{nonpert}) was one of the cases studied.} For example a
minimum with $g_{\STR}^{2} = 1/2$ can be found for the choice of
parameters
\begin{equation}
A_0 = 8.9 \qquad A_1 = -4.5 \qquad B= 0.75 . \label{parameters}
\end{equation}
The choice in~(\ref{parameters}) also has the pleasant feature
that $f(\lang \ell \rang) \approx 0$ so that from~(\ref{lingauge})
we have $\lang \ell \rang \approx g_{\STR}^{2}/2$ as it would be
in the perturbative limit.

The explicit model of~\cite{BiGaWu96,BiGaWu97a} incorporates this
K\"ahler stabilization mechanism with a realistic model of gaugino
condensation in the hidden sector which includes modular
invariance, possible string threshold corrections to gauge
couplings as well as possible matter condensates in the hidden
sector. Yet despite these many complications, the dilaton
dependence of the condensate-induced superpotential, when written
in terms of the chiral formulation, continues to be of the form
$W(S) \sim e^{-3S/2b_{a}}$. Thus it should provide a manifestation
of~(\ref{Ktrue}), and indeed, upon translating from the linear
multiplet to the chiral multiplet notation using
\begin{equation}
\lang K_s \rang = -\ell \qquad \qquad \lang K_{s\bar{s}}\rang =
\frac{\ell^2}{1+\ell g'(\ell)}  \label{translate}
\end{equation}
that is exactly what happens, as was shown in~\cite{BiGaNe01}.

\subsection{Soft terms and benchmark choices}
\label{sec:soft1}

Implementing the ideas of Section~\ref{sec:motiv1}, in conjunction
with an explicit model of supersymmetry breaking via gaugino
condensation as in Section~\ref{sec:concrete1}, has the power to
relate parameters of the hidden sector to the scale of gaugino
condensation and the size of the gravitino mass, thereby providing
a complete model with a great deal of predictability. We need not
concern ourselves with such model-dependent issues here. The
discussion in Section~\ref{sec:concrete1} is meant merely to
illustrate the degree to which the scenario we are describing is
motivated by honest, semi-realistic models from string effective
field theory. For the purposes of generating benchmark scenarios
we can treat the gravitino mass $m_{3/2}$ and the beta-function
coefficient $b_{+}$ as independent parameters -- or even more
phenomenologically, treat the gravitino mass and $a_{\rm np}$
of~(\ref{acond}) as free parameters -- and investigate what sort
of departures from the standard phenomenology of the cMSSM we
might expect.

As mentioned previously, the impact of the K\"ahler suppression
factor $a_{\rm np}$ will be maximized when the dilaton is the sole
participant in supersymmetry breaking, as is in fact the case in
the explicit model of~\cite{BiGaWu97a}. From~(\ref{FS}) we see
that
\begin{equation}
\left| \frac{F^{S}}{M} \right| = \left| \frac{F^{S}}{3m_{3/2}}
\right| \simeq \frac{4a_{\rm np}}{\sqrt{3}} \ll 1
\end{equation}
so one-loop corrections can be important for those soft
supersymmetry-breaking terms that receive their tree level
contributions solely from the dilaton auxiliary field, such as the
gaugino masses and trilinear A-terms~\cite{BiGaWu97b}. In
particular, loop-corrections arising from the conformal anomaly
are proportional to $M$ itself and receive no suppression, so they
can be competitive with the tree level contributions in the
presence of a nontrivial K\"ahler potential for the dilaton and
should be included~\cite{GaNeWu99,BiGaNe01}. If we assume that the
K\"ahler metric for the observable sector matter fields is
independent of the dilaton, as is the case at tree level in
orbifold compactifications, then the leading order expressions for
the soft supersymmetry-breaking terms for canonically normalized
fields are
\begin{eqnarray}
M_{a} &\simeq& \frac{g_{a}^{2}(\mu)}{2}\[\lang F^S \rang
-2b_{a}m_{3/2}\] \nonumber \\ A_{ijk} &\simeq& -\lang K_s F^{S}
\rang + m_{3/2}\[\gamma_{i} + \gamma_{j} +\gamma_{k} \] \nonumber
\\ m_{0}^{2} &\simeq& m_{3/2}^{2} ,\label{softterms1}
\end{eqnarray}
where $\gamma_{i}$ is the anomalous dimension of field $Z^{i}$.
Complete expressions for these soft terms, as well as a brief
description of how soft terms are derived from string theory more
generally, are given in the Appendix. In the above expressions we
have made a tacit choice of relative phase between terms involving
$\lang F^S \rang$ and those involving $\lang M\rang =-3m_{3/2}$
such that the combination of terms will reduce $M_3$ and enhance
$M_1$ and $M_2$. We will also take $\lang K_s \rang =
-g_{\STR}^{2}/2$ and thus assume that the tree-level relationship
between the dilaton and the coupling constant is not affected
greatly by the presence of the nonperturbative corrections to the
dilaton K\"ahler potential. While we have presented only the
leading terms in the one-loop parameters in~(\ref{softterms1}),
the complete expressions for soft terms at one loop will be used
in our calculations.

From~(\ref{softterms1}) it is clear that the dominant signature of
a ``generalized'' dilaton-domination scenario is the hierarchy
between gaugino and scalar masses, as was noticed by
Casas~\cite{Ca96}. Indeed, comparing the first (dilaton-dependent)
term in the gaugino mass of~(\ref{softterms1}) to the scalar mass,
and using~(\ref{FS}) we have, for properly normalized fields at
tree level, the ratio
\begin{equation}
\left| \frac{M_{a}}{m_{0}} \right| = \sqrt{3} a_{\rm np}
\frac{g_{a}^{2} (\Lambda_{\UV})}{g_{\STR}^{2}} \label{ratio1}
\end{equation}
which reduces to the familiar factor of $\sqrt{3}$ of minimal
supergravity in the perturbative case $a_{\rm np} =1$ when the
boundary condition scale $\Lambda_{\UV}$ and the GUT scale are
taken to coincide. However, if we imagine the value of $a_{\rm
np}$ to be determined by the beta-function coefficient of a
hidden-sector gauge group as in~(\ref{aBGW}), then the {\em
largest} it can be is $a_{\rm np} \simeq 1/6.1$ which occurs when
the condensing gauge group is $E_8$. For the more realistic case
of a smaller condensing group the value of $a_{\rm np}$ will be
smaller, and hierarchies of $\order(10)$ between the scalar masses
and gaugino masses are common.

On top of this gross feature it is also clear that the loop
effects will produce a ``fine-structure'' of nonuniversalities
among the gaugino masses and A-terms. With the phase choice
represented in~(\ref{softterms1}) and the definition~(\ref{ba}) it
is clear that the effect of the loop corrections will be to lower
the gluino mass $M_3$ while increasing the bino mass $M_1$
relative to the wino $M_2$. In fact, for small enough $a_{\rm np}$
(or, equivalently, small enough $b_{+}$) it is possible to so
suppress the universal tree level contributions to the gaugino
masses and A-terms that the anomaly-mediated terms dominate and we
encounter a gaugino sector identical to that of the
anomaly-mediated supersymmetry breaking (AMSB) scenario with its
wino-like LSP~\cite{GiLuMuRa98,RaSu99,PoRa99}, only with large
(and positive) scalar masses for all matter fields. In general,
though, the ``anomaly-mediated'' terms (proportional to the
auxiliary field $M$ of supergravity) and the standard
``gravity-mediated'' terms (proportional to the auxiliary field
$F^S$ for the dilaton) will be comparable.\footnote{Strictly
speaking, the terms that have come to be referred to as
``anomaly-mediated,'' and indeed the whole paradigm that is
referred to as ``anomaly-mediated supersymmetry breaking,'' is
really a special case of gravity-mediation.}
It is important to note that the significant splitting experienced
by the gaugino masses is not also seen in the gauge couplings
themselves. Tree level gaugino masses are still universal, but are
suppressed, so that nonuniversal loop contributions are
comparable, while loop contributions to the gauge couplings
themselves are always small in comparison to the large tree level
value.

We will choose three points in the parameter space determined by
$\lbr \tan\beta, m_{3/2}, a_{\rm np} \rbr$ for further study in
Section~\ref{sec:phenom}. Note that this parameter set is meant to
replace those of the usual minimal supergravity, or cMSSM,
parameter set. We begin by setting the initial input scale to be
the GUT scale $\Lambda_{\UV} = 2 \times 10^{16} \GeV$ as this is a
common convention in the literature and makes for easier
comparisons with previous results. The phenomenology of this class
of models was studied at some length in~\cite{GaNe00a} where it
was found that requiring $m_{3/2} \approx 1 \TeV$ to within an
order of magnitude typically required $b_{+} \leq 0.15$. This is
consistent with recent studies of the hidden sector in realistic
compactification of heterotic string theory on $Z_3$
orbifolds~\cite{Gi01b,Gi02} where hidden sector gauge groups
larger than $SU(5)$ were very rare. We are thus led to consider
among our benchmark points the cases where $b_+ = 15/16\pi^2
\simeq 0.095$ and $b_+ = 9/16\pi^2 \simeq 0.057$. The former could
result from a condensation of pure $SU(5)$ Yang-Mills fields in
the hidden sector. The latter case could be obtained either from a
similar condensation of pure $SU(3)$ Yang-Mills fields or from the
condensation of an $E_6$ hidden sector gauge group with 9
$\mathbf{27}$'s condensing in the hidden sector as well. To serve
as a baseline, we will also consider a much larger value of the
condensing group beta-function coefficient of $b_{+}= 36/16\pi^2
\simeq 0.228$. This could result from a hidden sector condensation
of pure $E_6$ Yang-Mills fields. Thus we will define our first
three benchmark points as follows:
\begin{eqnarray}
{\rm Case\; A:} \qquad  \lbr \tan\beta,\; m_{3/2},\; a_{\rm np}
\rbr & = & \lbr 10,\; 1500 \GeV,\;  1/15.77 \rbr \label{benchA} \\
{\rm Case\; B:} \qquad \lbr \tan\beta,\; m_{3/2},\; a_{\rm np}
\rbr & = & \lbr 5,\; 3200 \GeV,\;  1/37.05 \rbr  \label{benchB} \\
{\rm Case\; C:} \qquad \lbr \tan\beta,\; m_{3/2},\; a_{\rm np}
\rbr & = & \lbr 5,\; 4300 \GeV,\;  1/61.36 \rbr . \label{benchC}
\end{eqnarray}
The corresponding values of the soft terms will be collected with
our other benchmark points in Table~\ref{tbl:inputs} in
Section~\ref{sec:phenom}. We have chosen to be precise in our
definitions of the parameter $a_{\rm np}$ so that the numbers in
Table~\ref{tbl:inputs} can be reproduced from the master equations
in the Appendix.

%% file: theory2g.tex
\subsection{Theoretical motivation}
\label{sec:motiv2}

In Section~\ref{sec:theory1} we considered the case where only the
dilaton participates in supersymmetry breaking. We now turn our
attention to the opposite case where it is only the K\"ahler
moduli (which are typically denoted by $T$) which communicate
supersymmetry breaking to the observable sector. This was found to
be a generic property of many early models of gaugino condensation
that used the tree-level form of the dilaton K\"ahler potential,
particularly those that employ multiple condensates to stabilize
the dilaton~\cite{CaLaMuRo90,deCaMu93a,deCaMu93b,deRo93}.

In the previous section we employed K\"ahler potential
stabilization of the dilaton to reduce the universal tree level
contribution to gaugino masses. In such a paradigm the dominant
loop corrections are those from the superconformal anomaly which
(depending on the relative phase of the dilaton auxiliary field
$F^S$ and the supergravity auxiliary field $M$) can reduce the
gluino mass at the boundary condition scale relative to the other
gaugino masses. When the dilaton plays no role in supersymmetry
breaking, however, the gaugino masses are entirely determined at
the loop level. Among these loop-level contributions there is a
universal contribution to gaugino masses in the form of the
universal Green-Schwarz counterterm, inherited from the underlying
string theory. For certain quite reasonable ranges for the
coefficient $\delta_{\GS}$ of this counterterm we will find that
the loop-induced gaugino mass arising from the $T$-moduli and the
superconformal anomaly are naturally comparable to the universal
term, leading to a lighter gluino than in the typical unified case
and diminished fine-tuning.

All orbifold compactifications of the weakly coupled heterotic
string give rise to certain chiral superfields which parameterize
the size and shape of the compact extra dimensions. It is always
possible to consider only a reduced set of three diagonal moduli
which we will denote $T^{\alpha}$, whose K\"ahler potential is
given by $K=-\sum_{\alpha} \ln(T^{\alpha}+\oline{T}^{\alpha})$.
With only a slight loss of generality in what follows we can
further simplify things by treating all $T^{\alpha}$ as equivalent
and thus $K=-3\ln(T+\oline{T})$.

The diagonal modular transformations
\begin{equation} T
\to \frac{aT - ib}{icT +d}, \;\;\; \; ad-bc=1, \; \;a,b,c,d \in Z,
\label{modtrans}
\end{equation}
leave the classical effective supergravity theory invariant,
though at the quantum level these transformations are
anomalous~\cite{DiKaLo91,AnNaTa91,CaOv93,DeFeKoZw92}. This anomaly
is cancelled in the effective theory by the presence of a
universal Green-Schwarz counterterm and model-dependent string
threshold corrections to the gauge kinetic
functions~\cite{GaTa92,KaLo95}, which we will have occasion to
describe below. A matter field $Z^{i}$ is said to have modular
weight $n_{i}$ if it transforms under~(\ref{modtrans}) as
\begin{equation} Z^i \to
(icT + d)^{n_i} Z^i . \label{mattertrans}
\end{equation}
In what follows we will assume that the matter fields universally
have modular weight $n_i = -1$, as would be the case for fields
arising from the untwisted sector of the heterotic string. This
will simplify the analysis of the gaugino masses. Such models have
often be referred to as orbifold models of "Type II", or O-II
Models, in the literature~\cite{BrIbMu94}.

Since the K\"ahler potential for matter fields, derived from the
tree level string theory, is given by the diagonal metric
\begin{equation} K_{i \bar{j}} = \kappa_i(Z^n) \delta_{ij} +
O(|Z^i|^2), \qquad  \kappa_i(Z^n) = (T + \oline{T})^{n_{i}} \to (T
+ \oline{T})^{-1} , \label{kappa}
\end{equation}
we see that~(\ref{modtrans}) is manifested as a K\"ahler
transformation $K \to K + 3(F+\oline{F})$, with $F=\ln(icT+d)$ and
the classical symmetry of the effective Lagrangian will be
preserved since under~(\ref{mattertrans}) with $n_i = -1$ the
superpotential transforms as~\cite{FeLuShTh89,FeLuTh89}
\begin{equation}
W \to W \(icT + d\)^{-3} = We^{-3F} . \label{pottrans}
\end{equation}

The above transformations are known to be preserved by the
underlying string theory to all orders in perturbation theory, and
are conjectured to hold even in the presence of nonperturbative
effects. Therefore the quantum level anomaly in the effective
theory must be cancelled by an appropriate set of operators. This
is provided in part by a Green-Schwarz counterterm with universal
coefficient $\delta_{\GS}$, which can be thought of as a
loop-correction (or genus one from the point of view of string
theory) that contributes to gaugino masses in the
form\footnote{Expressions for soft terms are understood from here
onwards as being taken as functions of the vacuum expectation
values of the fields involved. Thus $s=\lang S \lowest \rang$,
$t=\lang T \lowest \rang$, etc.}
\begin{equation} M^{1}_a|_{\GS}
= \frac{g^2_a(\mu)}{2}\frac{2F^T}{(t + \bar{t})}
\frac{\delta_{\GS}}{16\pi^2} . \label{MGS}
\end{equation}
Here $\delta_{\GS}$ is a (negative) integer which is calculable
from the string compactification and whose value ranges from 0 to
-90 in the normalization adopted in~(\ref{MGS}). When $\lang F^T
\rang \neq 0$ this term provides a universal contribution to
gaugino masses at the loop level.

Let us next examine the issue of dilaton stabilization in this
class of models. Taking the simplest case of just two gaugino
condensates in the hidden sector we would expect a superpotential
of the form
\begin{equation}
W(S,T) = \Lambda_{\STR}^{3} f(T) \[d_{1}e^{-3S/2b_{1}} +
d_{2}e^{-3S/2b_{2}}\] , \label{raceW}
\end{equation}
where $f(T)$ is a function of the moduli $T$ which depends on the
value of the Green-Schwarz coefficient $\delta_{\GS}$.
In~(\ref{raceW}) $b_{1}$ and $b_{2}$ are the beta-function
coefficients, defined by~(\ref{ba}), for the two condensing groups
${\cal G}_{1}$ and ${\cal G}_{2}$ and $d_{1}$ and $d_{2}$
parameterize the presence of possible matter in the hidden sector.

For dilaton stabilization to occur, we must require that the
scalar potential $V(S)$ given in~(\ref{dilpot}) give rise to a
minimum such that $\lang s+\bar{s}\rang/2 = 1/g_{\STR}^{2} \simeq
2$ while generating a gravitino mass $m_{3/2} = \lang e^{K/2}
W\rang$ of $\order(1 \TeV)$. Here we will no longer require the
presence of nonperturbative corrections to the dilaton K\"ahler
potential so that $K_{s\bar{s}}^{1/2}=1/(S+\bar{S})$, {\em i.e.}
$a_{\rm np} = 1$. Then minimizing the dilaton potential
with~(\ref{raceW}) yields the vacuum solutions\footnote{These
solutions are strictly true only in the limiting case
$\delta_{\GS} \to 0$ when the chiral formulation is used for the
dilaton. However, as we will be considering relatively small
values for this coefficient in Section~\ref{sec:concrete2} this
approximation is justified. For more details,
see~\cite{deCaMu93a,deCaMu93b}.}
\begin{eqnarray}
{\rm Im}\; s&=& \frac{2\pi (2n+1)}{3(b_{1}^{-1} - b_{2}^{-1})}
\qquad n \in Z \nonumber \\ {\rm Re}\; s & = &
\frac{2}{3(b_{1}^{-1} - b_{2}^{-1})}\ln\[\frac{d_{1}(3 (\re \; s)
b_{1}^{-1} + 1)}{d_{2}(3 (\re \; s) b_{2}^{-1} + 1)}\] .
\label{raceSOL}
\end{eqnarray}
The first of these equations merely introduces a relative sign
between the two condensates to produce a minimum. As for the
second equation in~(\ref{raceSOL}), if we succeed in achieving a
realistic vacuum solution we expect $3(\re \;s)/b_{a} \gg 1$ for
both condensing groups ${\cal G}_{a}$. In that case we can
approximate the second solution as
\begin{equation}
{\rm Re}\; s \simeq \frac{2}{3(b_{1}^{-1} -
b_{2}^{-1})}\ln\frac{d_{1}b_{2}}{d_{2}b_{1}} . \label{raceS}
\end{equation}
Each of these four parameters $b_{1}$, $b_{2}$, $d_{1}$ and
$d_{2}$ are not continuously variable, but depend upon the gauge
groups in the hidden sector and the representation and number of
fields in the hidden sector charged under those groups. They are
calculable, however, from any particular orbifold compactification
of the weakly coupled heterotic string.

\subsection{A concrete realization}
\label{sec:concrete2}

In this section we are
following~\cite{CaLaMuRo90,deCaMu93a,deCaMu93b,deRo93} in working
with the tree level K\"ahler potential for the dilaton, modified
only by the presence of a Green-Schwarz counterterm. This will
lead to a minimum where $\lang F^S \rang = 0$ is favored while
$\lang F^T \rang \neq 0$ to provide for supersymmetry
breaking~\cite{FoIbLuQu90,CvFoIbLuQu91}. In these cases it was
found that under a variety of different forms for $f(T)$ in the
condensate superpotential the $T$ moduli are stabilized at $\lang
\re \; t  \rang \simeq 1.2$ which implies that the compact space
has a typical dimension of order the inverse Planck mass.

An exact solution that generates both $m_{3/2}$ and $\lang \re\;
s\rang$ requires a model for the coefficients $d_{1}$ and $d_{2}$
in~(\ref{raceS}). Several possibilities for generating differing
values of these coefficients exist in the literature.
In~\cite{CaLaMuRo90} these coefficients represented threshold
effects in the beta functions for the couplings of groups ${\cal
G}_{1}$ and ${\cal G}_{2}$, due to the integrating out of heavy
vector-like matter charged under those groups, with masses above
the supersymmetry breaking scale. If the hidden sector matter is
to be integrated out below the scale of gaugino condensation then
a nontrivial $d_{a}$ is generated for each condensing group whose
form depends on whether the matter is vector-like in
nature~\cite{deCaMu93a} or only forms condensates of dimension
three or higher~\cite{BiGaWu97a}.

Literally hundreds of explicit examples where~(\ref{raceS})
produced a minimum at $g_{\STR}^{2} \simeq 1/2$ and $m_{3/2}$
within an order of magnitude of $1 \TeV$ were obtained
in~\cite{deCaMu93a}, both with and without a nonzero Green-Schwarz
coefficient. These cases involved taking ${\cal G}_{1} = SU(N_1)$
and ${\cal G}_{2} = SU(N_2)$ with differing numbers of
fundamentals charged under each group. For example, taking the
limit $\delta_{\GS} = 0$ for the moment, a hidden sector
comprising of ${\cal G}_{1} = SU(7)$ with 8
$\mathbf{7}+\mathbf{\bar{7}}$'s and ${\cal G}_{2} = SU(8)$ with 15
$\mathbf{8}+\mathbf{\bar{8}}$'s satisfied~(\ref{raceS}) with
$\lang \re \; s\rang = 2.0$ and $m_{3/2} = 2550 \GeV$. In another
example with somewhat stronger string coupling the hidden sector
was given by ${\cal G}_{1} = SU(5)$ with 11
$\mathbf{5}+\mathbf{\bar{5}}$'s and ${\cal G}_{2} = SU(7)$ with 3
$\mathbf{7}+\mathbf{\bar{7}}$'s satisfied~(\ref{raceS}) with
$\lang \re \; s\rang = 1.2$ and $m_{3/2} = 6526 \GeV$. Lastly, an
example with slightly weaker string coupling had a hidden sector
of pure ${\cal G}_{1} = SU(7)$ Yang-Mills fields and ${\cal G}_{2}
= SU(8)$ with 7 $\mathbf{8}+\mathbf{\bar{8}}$'s that yielded a
gravitino mass of $m_{3/2} = 32 \TeV$ and $\lang \re \; s \rang =
2.2$. In short, so many possible combinations of gauge groups with
the desired properties have been catalogued that we will feel
justified in what follows to assume $g_{\STR} \simeq 1/2$ while
treating $m_{3/2}$ as a free parameter of the theory. This will
allow us the freedom to study this entire class of models without
appealing to specific constructions of the hidden sector.

Finally, as for the negative integer value of the Green-Schwarz
coefficient $\delta_{\GS}$, this can be computed for any given
orbifold compactification. The simplified moduli sector we are
considering here is suggested by the phenomenologically
well-motivated $Z_3$ orbifold. The value of $\delta_{\GS}$ for all
such orbifold compactifications which could potentially give rise
to the Standard Model in the observable sector was recently
carried out~\cite{Gi01b}, in which it was found that the range of
possible values was actually quite limited and given by the set
\begin{equation}
\delta_{\GS} \in \lbr -9, -12, -15, -18, -24 \rbr . \label{dGS}
\end{equation}
Remarkably, we will find that the parameter combination $\lang \re
\; t \rang \gappeq 1$ and $9 \leq |\delta_{\GS}| \leq 24$ are
precisely the ranges that give rise to a light gluino which may be
produced at the Tevatron and which can ameliorate the fine-tuning
problems of the electroweak sector.

\subsection{Soft terms and benchmark choices} \label{sec:soft2}

Typically, the multiple condensate models with tree level dilaton
K\"ahler potential are incapable of achieving a vanishing vacuum
energy at the minimum of the scalar
potential~\cite{CaLaMuRo90,deCaMu93a,Ca96}. However, without
ensuring $\lang V \rang = 0$ it is unclear whether a meaningful
analysis of low-energy phenomenology is possible (see, for
example, the discussion of this point in~\cite{BrIbMu94}). Rather
than introduce additional model dependence by incorporating a new
sector in the theory to cancel the residual vacuum energy we will
instead assume some implicit mechanism that results in the
vanishing of the potential~(\ref{pot}) at its
minimum~\cite{BaLo99}. This allows us to determine the vev of the
auxiliary field $F^T$ in the moduli-dominated limit as $\lang F^T
\rang = m_{3/2}\lang(t+\bar{t})\rang$. Then the gaugino sector is
determined by the three parameters $m_{3/2}$, $\delta_{\GS}$ and
$\lang \re \;t \rang$.\footnote{Since we have dropped phases in
the gaugino masses we will consider only real values of the
overall $T$ modulus.}

In the moduli-dominated limit $\lang F^S \rang = 0$, with $n_{i} =
-1$ for all observable sector fields, we obtain for the full
one-loop gaugino masses
\begin{equation}
M_{a} = \frac{g_{a}^{2}\(\Lambda_{\STR}\)}{2} \lbr 2
 \[ \frac{\delta_{\GS}}{16\pi^{2}} + b_{a}
\]\Eisen F^{T} + \frac{2}{3}b_{a}\oline{M} \rbr , \label{Maloop}
\end{equation}
where we have introduced the modified Eisenstein function
\begin{equation}
\Eisen \equiv \left(2\zeta(t) + \frac{1}{t+\bar{t}}\right)
\label{Eisenstein}
\end{equation}
with the Riemann zeta function and classical Dedekind function
$\eta(T)$ given by
\begin{equation}
\eta(T) = e^{-\pi T /12} \prod_{n=1}^{\infty} (1-e^{-2\pi nT});
\qquad \zeta(T) = \frac{1}{\eta(T)} \frac{d\eta(T)}{dT} .
\end{equation}
For our purposes we need only bear in mind that the modified
Eisenstein function~(\ref{Eisenstein}) vanishes at the self-dual
points $<t>\;=1$ and $<t>\;=e^{i\pi/6}$.

\begin{figure}[t]
    \begin{center}
\centerline{
       \epsfig{file=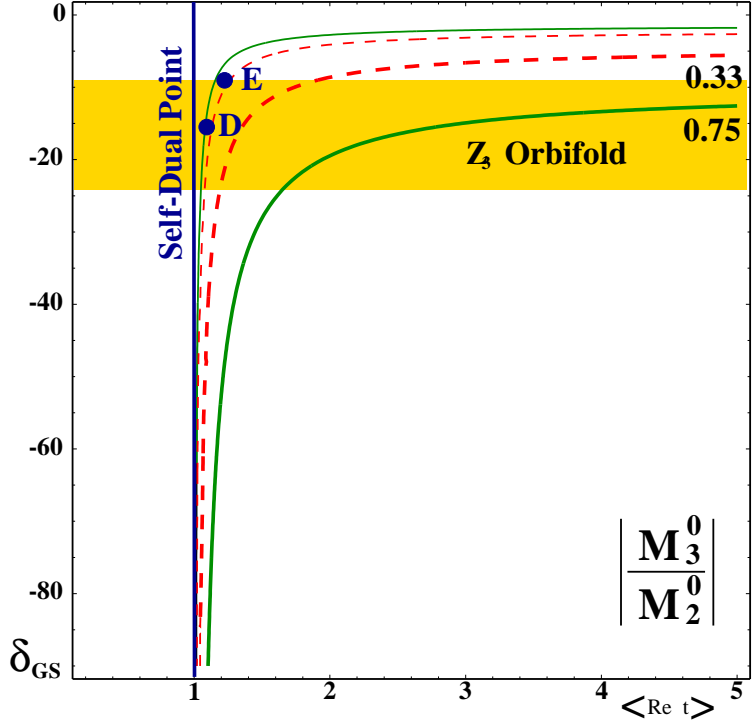,width=0.6\textwidth}}
          \caption{\footnotesize{\label{fig:modgluino} {\bf Ratio of unification scale gluino mass to wino mass as a function of
          Green-Schwarz coefficient $\delta_{\GS}$ and
          $\lang \re \; t \rang \geq 1$}. Contours of $|M_3/M_2|$ at the boundary condition scale
          of $\Lambda_{\UV} = 2 \times 10^{16} \GeV$ of 0.75 (solid green)
          and 0.33 (dashed red) are given. The upper set of contours have ${\rm sgn}(M_3)
          =-{\rm sgn}(M_2)$, while for the lower set of contours ${\rm sgn}(M_3)
          ={\rm sgn}(M_2)$. The preferred region indicated by~(\ref{dGS})
          for the $Z_3$ orbifold is shown by the shaded region. We have indicated
          the position of our two benchmark points~D and~E.}}
    \end{center}
\end{figure}

In~(\ref{Maloop}) we see the essential elements for addressing the
fine-tuning in the electroweak sector: A universal contribution
from the Green-Schwarz counterterm and group-dependent
contributions which distinguish the gauginos of the asymptotically
free $SU(3)$ group from the others. The interplay between these
contributions, without the anomaly contribution in~(\ref{Maloop}),
was studied in orbifold models in~\cite{ChDrGu97}. For the right
combinations of relative phase between $F^T$ and $M$ (we have
tacitly assumed zero relative phase) and sign of $\Eisen$ it is
possible to diminish the gluino mass relative to the other
gauginos. This is exhibited in Figure~\ref{fig:modgluino} where we
have highlighted the region preferred by the $Z_3$ orbifold.
Remarkably, it appears that the $Z_3$ orbifold, with moduli
stabilized just slightly away from their self-dual points,
actually prefers a light gluino.

To complete our model and generate spectra for use as benchmarks
we need to exhibit the remainder of the soft supersymmetry
breaking terms. The insistence on modular weights $n_i = -1$
generates a model similar to the no-scale models in which all soft
terms are zero at the tree level, independent of the ultimate
value of $\lang t \rang $ provided $\lang F^S \rang = 0$. As a
result, much of the one loop correction to the various soft terms
will also vanish, as they are proportional to tree level soft
supersymmetry breaking. The remainder of the one loop soft term
contributions depend on the manner in which the theory is
regulated. The complete set of one  loop terms was computed in
full generality in~\cite{GaNe00b} and specialized to the cases we
are considering here in~\cite{BiGaNe01}. We reserve further
details for the Appendix. The full set of soft supersymmetry
breaking terms we will employ are then
\begin{eqnarray}
M_{a}&=&\frac{g_{a}^{2}\(\mu\)}{2} \lbr 2
 \[ \frac{\delta_{\GS}}{16\pi^{2}} + b_{a}
\]\Eisen F^{T} + \frac{2}{3}b_{a}\oline{M} \rbr \nonumber \\
A_{ijk}&=&m_{3/2}\[\gamma_i + \gamma_j +\gamma_k\] \nonumber \\
m_{i}^{2}&=&\gamma_i m_{3/2}^{2}. \label{modterms}
\end{eqnarray}
Let us note that the scalar masses in~(\ref{modterms}) are truly
anomaly mediated, in the sense that their origin lies in the
superconformal anomaly and they are proportional to the
supergravity auxiliary field directly, though they are nonzero at
one loop and positive for the matter fields (though potentially
negative for the two Higgs fields of the MSSM, depending on the
value of $\tan\beta$). This is in contrast to the masses found
in~\cite{RaSu99,PoRa99} and subsequent work, which are a special
case of the more generalized anomaly-induced soft terms found
in~\cite{GaNe00b}. The case considered here differs from the
``standard'' AMSB in the assumptions made about the regularization
of the theory. We refrain from designating this true ``anomaly
mediation'' because the soft terms considered here are manifestly
not insensitive to UV physics, as is the hallmark of the anomaly
mediated supersymmetry breaking paradigm. But just as in the cases
of anomaly mediation so often considered in the literature, these
models will contain nearly-degenerate charginos and lightest
neutralinos and have a typical supersymmetry breaking scale of
$\order(10-20 \TeV)$. Given the discussion in
Section~\ref{sec:concrete2} we will define our second two
benchmark points then as follows:
\begin{eqnarray}
{\rm Case\; D:} \qquad \lbr \tan\beta,\; m_{3/2},\;
\delta_{\GS},\; \lang \re \; t\rang \rbr & = & \lbr 45,\; 20
\TeV,\; -15, \; 1.10 \rbr \label{benchD} \\ {\rm Case\; E:} \qquad
\lbr \tan\beta,\; m_{3/2}, \delta_{\GS},\; \lang \re \; t \rang
\rbr & = & \lbr 30,\; 20 \TeV,\; -9, \; 1.23 \rbr , \label{benchE}
\end{eqnarray}
with the corresponding values of the soft terms to be given in
Table~\ref{tbl:inputs} in Section~\ref{sec:phenom} below. We
choose to impose these boundary conditions at the scale
$\Lambda_{\UV} = 2 \times 10^{16} \GeV$ as in the previous
section.

%% file: theory3g.tex
\subsection{Theoretical motivation}
\label{sec:motiv3}

Models of gauge mediation~\cite{GiRa99} are typically
characterized by a scale at which supersymmetry breaking is
transmitted to the observable sector that is far lower than the
Planck or string scale. But this need not necessarily be the case.
Of course if one wants the soft supersymmetry breaking terms to be
{\em dominated} by the gauge-mediated contribution then one needs
to suppress the relative gravity-mediated contribution which is
always present. This can be accomplished simply by making the mass
scale of the messenger particles much smaller than the string
scale (the mass scale of the ``messengers'' in gravity-mediated
models). The idea of gauge mediation drew its greatest motivation
by the desire to have supersymmetry breaking communicated to the
Standard Model at an energy scale far below any possible scale of
flavor physics -- hence the tendency to demand mass scales on the
order of $100 \TeV$ for messenger fields and gravitino masses far
below $1 \GeV$.

Here we will not try to address the problems of flavor that may be
present in string-derived supergravity models, but merely address
the possibility that gauge mediation of supersymmetry breaking
from a hidden sector to the observable sector may well exist in
addition to the standard gravity-mediated mechanism. In fact,
given the generic occurrence of additional exotic vector-like
pairs of matter charged under observable sector gauge groups in
semi-realistic string
compactifications~\cite{Gi02,ChCoFa96,ClFaNa99,ClFaNa01,ClCvEsEvLaWa99}
we can conclude that ``partial'' gauge mediation most certainly
{\em does} occur in string-derived models. The only question is to
whether or not these contributions to soft terms are comparable in
size to those we described in the previous sections. In the
approaches we will examine they are naturally comparable. The idea
of combining gauge and gravity mediation is not new, particularly
in the context of string theory~\cite{AnBe92,Fa96a,Fa96b,deIr00}.

To review the basic elements of gauge mediation that we will need,
let us begin with the messenger sector. We imagine a set of chiral
fields $\Phi_i$ and $\oline{\Phi}_i$ that come in vector-like
representations of one or more of the subgroups of the Standard
Model. These fields experience a superpotential coupling to a
chiral field $X$, which is a singlet under the gauge groups of the
Standard Model, of
\begin{equation}
W = \lambda_{i} \oline{\Phi}_i X \Phi_i \label{messmass}
\end{equation}
so that should the chiral field $X$ receive a vacuum value $\lang
X \rang = M_{X}$ in its lowest component, we would have Dirac
fermions with masses $\sim \lambda_i M_{X}$. Note that we have
chosen a single field $X$ and diagonal couplings $\lambda_{ij} =
\delta_{ij}\lambda_i$ in~(\ref{messmass}) for
simplicity.\footnote{We are also tacitly assuming, in the spirit
of low-energy gauge-mediated models, that the K\"ahler potentials
for both the messengers and the singlet field $X$ are trivial.
This is not generally true in superstring constructions but we can
imagine absorbing the moduli dependence, such as the factors of
$\kappa_i$ in~(\ref{kappa}), into the vacuum values of $F^X$ and
$M_{X}$.} The index $i$ can be thought of as counting the number
of copies, or ``flavors,'' of each messenger field $\Phi$. The
field $X$ is further assumed to carry the information of
supersymmetry breaking through a nonzero highest component, so
that $\lang X \rang = M_{X} + \theta^2 F^X$, and thus the
messenger sector has mass splittings between its scalar and
fermionic components of order $\sqrt{F^X}$.

When the messenger mass scale $M_{X}$ is lower than the GUT scale
it is typical to employ messengers which form complete multiplets
under a unified group such as $SU(5)$. This ensures that gauge
coupling unification is preserved while providing a certain
universality in the soft term expressions. From a string theory
perspective it is preferable to relax this assumption, so we will
instead invoke incomplete GUT multiplets as
messengers~\cite{Ma97}, though we will continue to assume a
universal mass splitting $F^X$ and adopt the simplification of a
universal Yukawa coupling $\lambda_{i}=1$ in~(\ref{messmass}) and
hence a universal messenger mass $M_X$. Each of the specific cases
we will deal with below will be designed to ensure gauge coupling
unification despite the incomplete GUT representations. It is of
use to introduce the standard {\em messenger index}
\begin{equation}
N_a=\sum_{i=1}^{N_F} n_a^i
\label{index}
\end{equation}
where $n_a^i$ is the Dynkin index for the representation
$\mathbf{r}$ with flavor index $i$ under each Standard Model gauge
group ${\cal G}_{a}$. It is normalized with a GUT normalization so
that $n_a = 1$ for a pair of $SU(N)$ fundamentals and $n_1 =
(6/5)Y^2$ for a messenger pair with hypercharge $Y$. While we
adopt the $SU(5)$ GUT normalization on the hypercharges of our
messenger fields so as to make contact with the standard cases in
the literature, it is important to note that in realistic string
constructions there is no reason to assume that vector-like
messenger fields will have the same hypercharges as their Standard
Model analogs~\cite{Gi02,ChHoLy95,DiFaRu96,ClFaSa01}.

With these definitions, the gauge-mediated contributions to
gaugino masses are given by
\begin{equation}
\delta
M_a(\Lambda_{\UV})=\frac{g_a^{2}(\Lambda_{\UV})}{16\pi^2}N_a
\frac{F^X}{M_X} \label{gauginomass}
\end{equation}
while those of the scalar masses are
\begin{equation}
\delta m_A^{2}(\Lambda_{\UV})  =  2\sum_a N_a
C_a^{A}\(\frac{g_a^{2}(\Lambda_{\UV})}{16\pi^2}\)^{2}
\(\frac{F^X}{M_X}\)^2 \label{scalarmass}
\end{equation}
where $C_a^A$ is the standard quadratic Casimir for (SM) particle
$\Phi^A$ with $C_a^A = (N^2 -1)/2N$ for $SU(N)$ fundamentals and
$(3/5)Y^2$ for hypercharge (properly normalized to GUT
normalization). Since we imagine here only those cases for which
$F^X \ll M_{X}^{2}$ we can dispense with all but the leading terms
in the functions $f(x)$ and $g(x)$ of~\cite{GiRa99,Ma97}. We
propose to add the contributions in~(\ref{gauginomass})
and~(\ref{scalarmass}) to those of the supergravity contributions
described in the previous two sections. We should note
that~(\ref{gauginomass}) and~(\ref{scalarmass}) were computed in
the $\oline{DR}$ renormalization scheme appropriate to global
supersymmetry~\cite{GMSBcalc} but here we wish to employ them in
cases where the messenger masses will be much closer to the Planck
scale, suggesting a regularization scheme appropriate to
supergravity (such as Pauli-Villars) in called for. For the
purposes of obtaining benchmark scenarios, however, we will ignore
this technical, though potentially interesting, issue.

Standard analyses of gauge mediated supersymmetry breaking now
would proceed by treating both $F^X$ and the messenger mass
$M_{X}$ as free variables, with $F^X$ ultimately determined by
some model-dependent mechanism which ensures $\lang W_X \rang \neq
0$, since nonrenormalizable or Planck-suppressed operators are
discarded. In the presence of supergravity the auxiliary field
$F^X$ is determined by
\begin{equation}
F^X = -e^{K/2M_{\PL}^{2}}\(\oline{W}_{\oline{X}} +
\frac{X\oline{W}}{M_{\PL}^{2}}\) ,
\label{FXlocal}
\end{equation}
where we have restored the reduced Planck mass $M_{\PL} = 2.4
\times 10^{18} \GeV$ for clarity. There is always a contribution,
independent of any additional superpotential terms involving $X$,
given by $F^X = - M_{X} m_{3/2}$ from the second term
in~(\ref{FXlocal}). But in supergravity theories the mass
splitting within the messenger sector is no longer given simply by
$F_X$, which must be replaced in~(\ref{gauginomass})
and~(\ref{scalarmass}) by the off-diagonal mass terms of the
complete supergravity potential.

For example in the minimal case $F^X = - M_{X} m_{3/2}$ ({\em
i.e.} $\lang W_X \rang \simeq 0$) then the messenger mass spectrum
would be determined by~(\ref{pot})
\begin{equation}
V\big|_{\rm SUGRA} \ni |\tilde{\lambda} X|^2 \(|\Phi|^2 +
|\oline{\Phi}|^2\) + m_{3/2}^2 \(|\Phi|^2 + |\oline{\Phi}|^2\) +
\(\tilde{\lambda} m_{3/2} M_X (M_X/M_{\PL})^2 \Phi \oline{\Phi} +
\hc \) , \label{messsectorLOCAL}
\end{equation}
where $\tilde{\lambda} = e^{K/2}\lambda$. Making the appropriate
substitutions for $F_X$ in~(\ref{gauginomass})
and~(\ref{scalarmass}) we can see that the gauge-mediated
contributions to soft terms in this case are proportional to the
gravitino mass with a typical size
\begin{equation}
m_{\rm soft} \sim m_{3/2}
\frac{1}{16\pi^2}\(\frac{M_X}{M_{\PL}}\)^2 . \label{minsoft}
\end{equation}
Even when tree level gravity-mediated soft terms are absent or
suppressed, as in Sections~\ref{sec:theory1}
and~\ref{sec:theory2}, such gauge-mediated contributions would
prove irrelevant unless the messenger mass $M_X$ was extremely
close to the Planck scale.

We are thus led to consider cases in which supersymmetry continues
to be broken in the hidden sector by one of the string moduli $S$
or $T$, generating an F-term of general magnitude $<F^0>\; \simeq
\sqrt{3}m_{3/2} M_{\PL}$ to bring about a vanishing vacuum energy
and generating a gravitino mass on the order of $1 \TeV$. In
addition we will allow for some undetermined mechanism to generate
a non-vanishing $\lang F_X \rang$ through the first term
in~(\ref{FXlocal}) as in typical gauge-mediated models. We will
parameterize the size of this additional source of supersymmetry
breaking through the parameter $k=F^X/F^0$, with $F^0$ identified
with either $F^S$ or $F^T$ as in the previous sections.

\subsection{A concrete realization}
\label{sec:concrete3}

One of the reasons that gauge mediation is not often considered in
the context of string theory is the difficulty in finding suitable
messenger sectors when only renormalizable couplings are allowed.
The fields need to be vector-like, charged under one or more of
the subgroups of the Standard Model, remain light down to very low
energies and be capable of communicating directly with the
supersymmetry breaking of the hidden sector through operators that
are not suppressed by powers of the Planck mass. Such
circumstances are rare in actual string compactifications. However
{\em any} massive vector-like pairs, charged under a subgroup of
the Standard Model, {\em can and will} participate in
gauge-mediation of supersymmetry breaking at least through the
supergravity-generated second term in~(\ref{FXlocal}). Since all
realistic string constructions contain such exotic vector-like
states we can assert that ``partial'' gauge-mediation is a generic
outcome of string theory that should not be neglected.

These potential messenger sectors tend to come in incomplete
multiplets of $SU(5)$, however, instead of the $\mathbf{5}$ and
$\oline{\mathbf{5}}$'s that are so commonly employed. What's more,
these models also predict an anomalous $U(1)$ whose breaking
occurs at a scale $\Lambda_X \sim 10^{16} - 10^{17} \GeV$. If the
singlet field $X$ were charged under this anomalous $U(1)$, as is
typical, then we might expect the messenger mass $\lang X \rang
\simeq M_{X}$ to be of this magnitude as well, providing a
concrete realization of the above scenario. Note that by assuming
the messenger mass scale to be at or near the GUT scale we will
preserve the apparent unification of gauge couplings in the MSSM
up to small corrections.

In the context of Section~\ref{sec:motiv3} we will allow $\lang
F^X \rang$ to be a free parameter due to some $\lang W_X \rang
\neq 0$. Now the gauge-mediated contributions can be as large or
larger than the tree-level supergravity contribution, depending on
the relative size of the messenger mass, so we will return to the
simplicity of the dilaton-dominated scenario of
Section~\ref{sec:theory1}. Taking the tree level K\"ahler
potential for the dilaton (no nonperturbative corrections so
$a_{\rm np} = 1$) and
\begin{equation}
\lang F^0 \rang = \lang F^S \rang =
\frac{2}{g_{\STR}^{2}}\sqrt{3}m_{3/2}M_{\PL} \label{FSnew}
\end{equation}
then we can employ~(\ref{gauginomass}) and~(\ref{scalarmass}) with
$\lang F^X\rang = k \lang F^S\rang$. Note that the gravity and
gauge-mediated contributions are competitive whenever
\begin{equation}
\frac{k}{16\pi^2}\frac{M_{\PL}}{M_{X}} \simeq 1 . \label{ratio}
\end{equation}
It is interesting to note that this equality is satisfied for
$k\simeq 1$ when the messenger mass is near the anomalous $U(1)_X$
scale.

For our messenger sector we will introduce $n_D$ pairs of
messengers which are triplets under $SU(3)$ and $n_L$ pairs of
messengers which are doublets under $SU(2)_L$. We will not
introduce any specific messengers which carry only Standard Model
hypercharge. In fact, we leave the hypercharge assignments of the
messenger fields a free variable and work only with the overall
messenger index $N_1 = \sum_{i} n_{i}^{i}$ which we treat as a
continuous free parameter. If our messenger fields happen to have
the hypercharge of their Standard Model analogs, then $N_{1} =
(1/5)(2 n_D + 3 n_L)$. We follow the standard practice of setting
the initial scale for our soft parameters at the messenger mass
scale $M_{X}$, leaving the free variables that define our models
as $\lbr \tan\beta,\; m_{3/2}, \; M_{X}, \; k, \; n_D, \; n_L, \;
N_1 \rbr$.

\subsection{Soft terms and benchmark choices}
\label{sec:soft3}

We seek cases where the gauge-mediated masses are of the same
order of magnitude as the gravitino mass, so we return to the case
with $\lang F^T \rang = 0$ and $\lang F^S \rang \neq 0$ with
$a_{\rm np} = 1$. Since we will only consider cases where $M_X
\simeq M_{\STR}$ let us take $g_1(M_X) \simeq g_2(M_X) \simeq
g_3(M_X) \simeq g_{\STR}$ so that the complete, properly
normalized, gaugino masses are given by
\begin{eqnarray}
M_3 &=& \sqrt{3}m_{3/2}\[1+
n_D\frac{k}{8\pi^2}\frac{M_{\PL}}{M_X}\] \nonumber \\ M_2 &=&
\sqrt{3}m_{3/2}\[1+ n_L\frac{k}{8\pi^2}\frac{M_{\PL}}{M_X}\]
\nonumber \\ M_1 &=& \sqrt{3}m_{3/2}\[1+
N_1\frac{k}{8\pi^2}\frac{M_{\PL}}{M_X}\] \label{gaugGMSB2}
\end{eqnarray}
and the remaining soft terms are
\begin{eqnarray}
m_Q^2 &=& m_{3/2}^{2}\[1+ 6\(\frac{4}{3}n_D+\frac{3}{4}n_L
+\frac{1}{60}N_1\)
\frac{k^2}{(8\pi^2)^2}\(\frac{M_{\PL}}{M_X}\)^{2}\] \nonumber \\
m_U^2 &=& m_{3/2}^{2}\[1+6\(\frac{4}{3}n_D + \frac{4}{15}N_1\)
\frac{k^2}{(8\pi^2)^2}\(\frac{M_{\PL}}{M_X}\)^{2}\] \nonumber \\
m_D^2 &=& m_{3/2}^{2}\[1+6\(\frac{4}{3}n_D + \frac{1}{15}N_1\)
\frac{k^2}{(8\pi^2)^2}\(\frac{M_{\PL}}{M_X}\)^{2}\] \nonumber \\
m_L^2 &=& m_{3/2}^{2}\[1+6\(\frac{3}{4}n_L + \frac{3}{20}N_1\)
\frac{k^2}{(8\pi^2)^2}\(\frac{M_{\PL}}{M_X}\)^{2}\] \nonumber \\
m_E^2 &=& m_{3/2}^{2}\[1+\frac{18}{5}N_1
\frac{k^2}{(8\pi^2)^2}\(\frac{M_{\PL}}{M_X}\)^{2}\] .
\label{scalarGMSB2} \\
\end{eqnarray}
with $m_{H_u}^{2} =m_{H_D}^{2} =m_L^2$ and $A=\sqrt{3}m_{3/2}$. We
have again chosen a positive relative sign between the
contributions to gaugino masses in~(\ref{gaugGMSB2}) from gauge
messengers and those arising from supergravity.

The choice of messenger indices is dictated by the need to obtain
sufficiently large radiative corrections to the lightest CP-even
Higgs mass. By introducing messengers charged solely under $SU(3)$
a heavy gluino is produced that can achieve the necessary Higgs
mass with light scalars. Such a scenario was considered in the
context of low-energy gauge mediation in the context of string
theory in~\cite{Fa96b}. When Standard Model-like hypercharge
assignments for these messengers are assumed the bino mass $M_1$
tends to be much larger than the wino mass $M_2$ at the initial
high-energy input scale, producing a gaugino sector similar to
that of anomaly-mediation. We have therefore chosen to allow
non-standard hypercharges for the messenger fields and have
selected a value for $N_1$ that gives gaugino mass in the
gravity-mediated regime. We take the messenger mass scale to be
intermediate between the GUT scale and the Planck scale:
$\Lambda_{\UV} = M_{X} = 8 \times 10^{16} \GeV$, which was a
typical anomalous $U(1)$ scale in those models that gave rise to
suitable messenger fields~\cite{Gi01b}. Our two benchmarks points
for this section are then given by the following parameter sets
\begin{equation}
{\rm Case\; F:} \qquad  \lbr \tan\beta,\; m_{3/2},\; n_D, \; n_L,
\; N_1 \rbr  =  \lbr 10,\; 120 \GeV,\;  4, \; 0, \; (3/5) \rbr
\label{benchG}
\end{equation}
\begin{equation}
{\rm Case\; G:} \qquad  \lbr \tan\beta,\; m_{3/2},\;  n_D, \; n_L,
\; N_1 \rbr  =  \lbr 20,\; 130 \GeV,\; 3, \; 0, \; (3/5) \rbr
\label{benchH}
\end{equation}
with $k=1$. Note that both of these examples are close in spirit
to the standard gauge-mediated examples in that the gravitino,
while not the LSP, is much lighter than the supergravity-dominated
models of our previous cases.

%% file: phenomg.tex
\subsection{Benchmark spectra and phenomenology}
\label{sec:spectra}

\begin{table}[th]
{\begin{footnotesize} {\begin{center}
\begin{tabular}{|l|c|c|c|c|c|c|c|} \cline{1-8}
Point & A & B & C & D & E & F & G \\  \cline{1-8} $\tan\beta$ & 10
& 5 & 5 & 45 & 30 & 10 & 20
\\ $\Lambda_{\UV}$ & $2 \times 10^{16}$ & $2 \times 10^{16}$ & $2 \times
10^{16}$ & $2 \times 10^{16}$ & $2 \times 10^{16}$ & $8 \times
10^{16}$ & $8 \times 10^{16}$
\\ \cline{1-8} $M_{1}$ & 198.7 & 220.1 & 215.3 & 606.5 & 710.8
& 278.9 & 302.2
\\ $M_{2}$ & 172.1 & 162.3 & 137.3 & 195.2 & 244.6 & 213.4 & 231.2
\\ $M_{3}$ & 154.6 & 122.3 & 82.4 & -99.2 & -89.0 & 525.4 & 482.9
\\ \cline{1-8} $A_t$ & 193.0 & 204.8 & 195.4 & 286.0 & 352.5 & 210.7 & 228.2
\\ $A_b$ & 205.3 & 235.3 & 236.3 & 390.6 & 501.5 & 211.6 & 229.2
\\ $A_{\tau}$ & 188.4 & 200.0 & 188.9 & 158.1 & 272.5 & 210.3 & 227.8
\\ \cline{1-8} $m_{Q_3}^{2}$ & $(1507)^2$ & $(3216)^2$ &
$(4323)^2$ & $(2035)^2$ & $(2144)^2$ & $(286)^2$ & $(276)^2$
\\ $m_{U_3}^{2}$ & $(1504)^2$ & $(3209)^2$ & $(4312)^2$ &
$(1487)^2$ & $(1601)^2$ & $(290)^2$ & $(281)^2$
\\ $m_{D_3}^{2}$ &  $(1505)^2$ & $(3213)^2$ & $(4319)^2$ & $(1713)^2$ &
$(1870)^2$ & $(287)^2$ & $(277)^2$
\\ $m_{L_3}^{2}$ & $(1503)^2$ & $(3208)^2$ & $(4312)^2$ & $(1361)^2$ &
$(1489)^2$ & $(125)^2$ & $(135)^2$ \\ $m_{E_3}^{2}$& $(1502)^2$ &
$(3206)^2$ & $(4308)^2$ & $(756)^2$ & $(1139)^2$ & $(140)^2$ &
$(152)^2$
\\ \cline{1-8} $m_{Q_{1,2}}^{2}$ & $(1508)^2$ & $(3220)^2$ & $(4328)^2$ &
$(2347)^2$& $(2347)^2$ & $(286)^2$ & $(276)^2$
\\ $m_{U_{1,2}}^{2}$ & $(1506)^2$ & $(3215)^2$ & $(4321)^2$ & $(2050)^2$
& $(2050)^2$ & $(290)^2$ & $(281)^2$ \\ $m_{D_{1,2}}^{2}$ &
$(1505)^2$ & $(3213)^2$ & $(4319)^2$ & $(1919)^2$ & $(1919)^2$ &
$(287)^2$ & $(277)^2$
\\ $m_{L_{1,2}}^{2}$ & $(1503)^2$ & $(3208)^2$ &
$(4312)^2$ & $(1533)^2$ & $(1533)^2$ & $(125)^2$ & $(135)^2$
\\ $m_{E_{1,2}}^{2}$ & $(1502)^2$ & $(3206)^2$ & $(4308)^2$ & $(1252)^2$
& $(1252)^2$ & $(140)^2$ & $(152)^2$   \\ \cline{1-8}
$m_{H_u}^{2}$ & $(1500)^2$ & $(3199)^2$ & $(4298)^2$ & $-(797)^2$
& $-(331)^2$ & $(125)^2$ & $(135)^2$
\\ $m_{H_d}^{2}$ & $(1503)^2$ & $(3208)^2$ & $(4312)^2$ & $(858)^2$
& $(1392)^2$ & $(125)^2$ & $(135)^2$ \\ \cline{1-8}
\end{tabular}
\end{center}}
\end{footnotesize}}
{\caption{\label{tbl:inputs}\footnotesize{\bf Soft Term Inputs}.
Initial values of supersymmetry breaking soft terms in GeV,
including the full one-loop contributions, at the initial scale
given by $\Lambda_{\UV}$. All points are taken to have $\mu > 0$.
The actual value of $\tan\beta$ is fixed in the EWSB conditions.
See the text for further discussion of the parameters and their
origins. Ways to convert these points into model lines are
discussed at the end of the Appendix. }}
\end{table}

The~\modelnum~benchmark scenarios described in
Sections~\ref{sec:theory1} through~\ref{sec:theory3} give rise
to~\modelnum~sets of high energy input values for renormalization
group (RG) evolution to the electroweak scale. These values are
determined by substituting the specified parameters into the
complete one-loop expressions for soft terms given in the
Appendix. The numerical value of these input quantities are
summarized in Table~\ref{tbl:inputs}. Actual evolution of these
parameters was carried out using the publicly-available code {\tt
SuSpect}~\cite{suspect} which performs RG integration at the
two-loop level from the specified input scale to the scale $M_z$.

{\tt SuSpect} uses the following quantities as inputs:
\begin{equation}
\alpha_{\EM}^{\oline{MS}}(M_Z)=1/127.938,\qquad
\alpha_s^{\oline{MS}}(M_Z)=0.118,\qquad\bar{s}^2_W=0.23117  ,
\label{SUSPECTalpha}
\end{equation}
as well as the following pole masses for heavy SM fermions:
\begin{equation}
M_t=174.3\GeV,\qquad M_b=4.62\GeV, \qquad M_\tau=1.778\GeV  .
\label{SUSPECTferm}
\end{equation}
Determination of parameters in the Higgs sector (such as the value
of the $\mu$ parameter inferred from EWSB) are computed at a scale
given by the geometric mean of the two stop masses. We have chosen
the number of iterations to achieve consistency in the EW sector
to be five. The light CP-even Higgs mass is calculated by using
the full one-loop tadpole method and includes leading NLO QCD
corrections as implemented in {\tt Subhpole}~\cite{Carena:1995wu}.
The values of $m_h$ determined by {\tt SuSpect} for the benchmark
scenarios have been checked against {\tt
FeynHiggs}~\cite{FeynHiggs} and found to be in acceptable
agreement. The radiative correction at NLO to all sparticles
masses are included and can be significant for many mass
eigenstates, particularly in the gaugino sector, affecting some
production rates and branching ratios noticeably. The resulting
low-energy spectrum for the~\modelnum~benchmark models is
summarized in Table~\ref{tbl:spectraSUM}.

\begin{table}[p]
{\begin{footnotesize} {\begin{center}
\begin{tabular}{|l|c|c|c|c|c|c|c|} \cline{1-8}
Point & A & B & C & D & E & F & G \\ \cline{1-8} $\tan\beta$ & 10
& 5 & 5 & 45 & 30 & 10 & 20
\\ $\Lambda_{\UV}$ & $2 \times 10^{16}$ & $2 \times 10^{16}$ & $2 \times
10^{16}$ & $2 \times 10^{16}$ & $2 \times 10^{16}$ & $8 \times
10^{16}$ & $8 \times 10^{16}$
\\ $m_{3/2}$ & 1500 & 3200 & 4300 & 20000 & 20000 & 120 & 130 \\
\cline{1-8} $M_1$ & 84.0 & 95.6 & 94.7 & 264.7 & 309.9 & 106.2 &
115.7
\\ $M_2$ & 133.7 & 127.9 & 108.9 & 159.0 & 198.5 & 154.6 & 169.6
\\ $M_3$ & 346.5 & 264.0 & 175.6 & -227.5 & -203.9 & 1201 & 1109 \\
\cline{1-8} $m_{\wtd{N}_1}$ & 77.9 & 93.1 & 90.6 & 171.6 & 213.0 &
103.5 & 113.1
\\ $m_{\wtd{N}_{2}}$ & 122.3 & 132.2 & 110.0 & 264.8 & 309.7 &
157.6 & 173.1
\\ $m_{\wtd{C}_{1}^{\pm}}$ & 119.8 & 131.9 & 109.8 & 171.6 & 213.0 & 157.5 & 173.0
\\ $m_{\tilde{g}}$ & 471 & 427 & 329 & 351 & 326 & 1252 & 1158
\\ $\wtd{B} \; \% |_{\rm LSP}$ & 89.8 \% & 98.7 \% &  93.4 \% &
 0 \% & 0 \% & 99.4 \% & 99.4 \%
\\ $\wtd{W}_{3} \% |_{\rm LSP}$ & 2.5 \% &  0.6 \% & 4.6 \% & 99.7 \% & 99.7 \% &
0.1 \% & 0.06 \%
\\ \cline{1-8} $m_{h}$ & 114.3 & 114.5 & 116.4 & 114.7 & 114.9 &
115.2 & 115.5
\\ $m_{A}$ & 1507 & 3318 & 4400 & 887 & 1792 & 721
& 640
\\ $m_{H}$ & 1510 & 3329 & 4417 & 916 & 1821 & 722 & 644 \\
$\mu$ & 245 & 631 & 481 & 1565 & 1542 & 703 & 643
\\ \cline{1-8} $m_{\tilde{t}_{1}}$ & 947 & 1909 & 2570 & 1066
& 1105 & 954 & 886 \\ $m_{\tilde{t}_{2}}$ & 1281 & 2639 & 3530 &
1678 & 1897 & 1123 & 991
  \\ $m_{\tilde{c}_{1}}$, $m_{\tilde{u}_{1}}$ & 1553 & 3254
& 4364 & 2085 & 2086 & 1127 & 1047
\\ $m_{\tilde{c}_{2}}$, $m_{\tilde{u}_{2}}$ & 1557 & 3260 & 4371 & 2382 &
2382 & 1132 & 1054
\\ \cline{1-8} $m_{\tilde{b}_{1}}$ & 1282 & 2681 & 3614 & 1213 &
1714 & 1053 & 971 \\ $m_{\tilde{b}_{2}}$ & 1540 & 3245 & 4353 &
1719 & 1921 & 1123 & 1037
\\ $m_{\tilde{s}_{1}}$, $m_{\tilde{d}_{1}}$ & 1552 & 3252 & 4362
& 1950 & 1948 & 1126 & 1045
\\ $m_{\tilde{s}_{2}}$, $m_{\tilde{d}_{2}}$ & 1560 & 3261 & 4372 & 2383 &
2384 & 1135 & 1057
\\ \cline{1-8} $m_{\tilde{\tau}_{1}}$ & 1491 & 3199 & 4298 & 559 & 1038
& 153 & 135
\\ $m_{\tilde{\tau}_{2}}$ & 1502 & 3207 & 4308 & 1321 & 1457 &
221 & 252 \\ $m_{\tilde{\mu}_{1}}$, $m_{\tilde{e}_{1}}$ & 1505 &
3207 & 4309 & 1274 & 1282 & 182 & 196
\\ $m_{\tilde{\mu}_{2}}$, $m_{\tilde{e}_{2}}$ & 1509 & 3211 & 4313 & 1544 &
1548 & 200 & 217 \\ $m_{\tilde{\nu_3}}$ & 1500 & 3206 & 4307 &
1314 & 1453 & 183 & 198  \\ \cline{1-8}
\end{tabular}
\end{center}}
\end{footnotesize}}
{\caption{\label{tbl:spectraSUM}\footnotesize {\bf Sample
Spectra}. All masses are in GeV.  For the purposes of calibrating
these results with those of other software packages we also
provide the running gaugino masses at the scale $M_Z$, which
include NLO corrections. See the text for further discussion of
the parameters and their origins. Ways to convert these points
into model lines are discussed at the end of the Appendix.}}
\end{table}

The benchmark models we present are quite interesting
phenomenologically, and should be studied in some detail. They
present challenges for present and future colliders that are
rather different from previously studied models. Here we will only
draw attention to a few general features.

\begin{figure}
\begin{center}
\includegraphics[scale=0.6,angle=270]{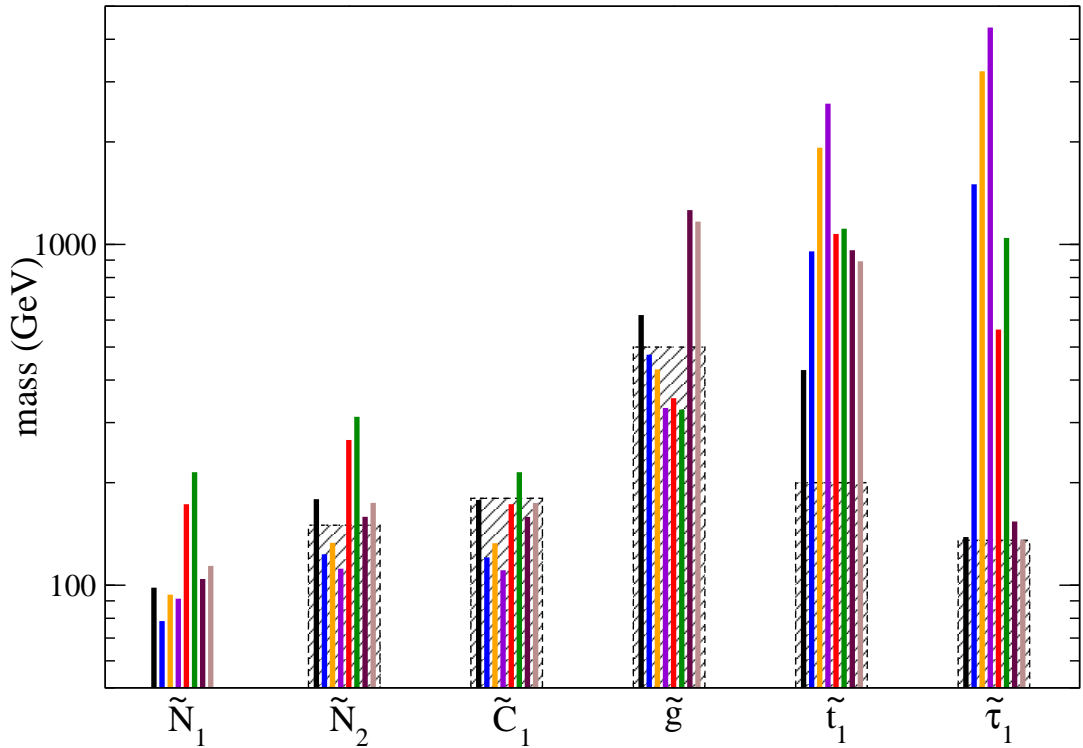}
\epsfxsize=15cm
\end{center}
{\caption{\label{fig:Tevspec}\footnotesize{\bf Sparticle masses
for different benchmark models.} For each superpartner, lines from
left to right correspond to mSUGRA Point~B of~\cite{bench}, and
our benchmark models A through G. The mass of the particle is
represented by the height of the line. Gray bars are crude
estimates of Tevatron reaches, {\em i.e.} any superpartner in the
gray region could be observed at the Tevatron. That reach of
course depends on backgrounds which will vary depending on the
details of the signal, so the gray regions are just approximate
guides for the non-expert reader. We have not studied the
detectability of these models at the LHC.}}
\end{figure}

Figure~\ref{fig:Tevspec} shows a summary of a number of the
superpartner masses, along with crude estimates of Tevatron
reaches. For comparison we also include the Snowmass benchmark
point most favorable for observation at the Tevatron~\cite{bench}.
In general the benchmark models we study have gauginos observable
at the Tevatron, as expected for supersymmetric worlds in which
electroweak symmetry breaking is explained by supersymmetry
without excessive fine tuning~\cite{KaLyNeWa02}.

\begin{figure}
\begin{center}
\includegraphics[scale=0.6,angle=270]{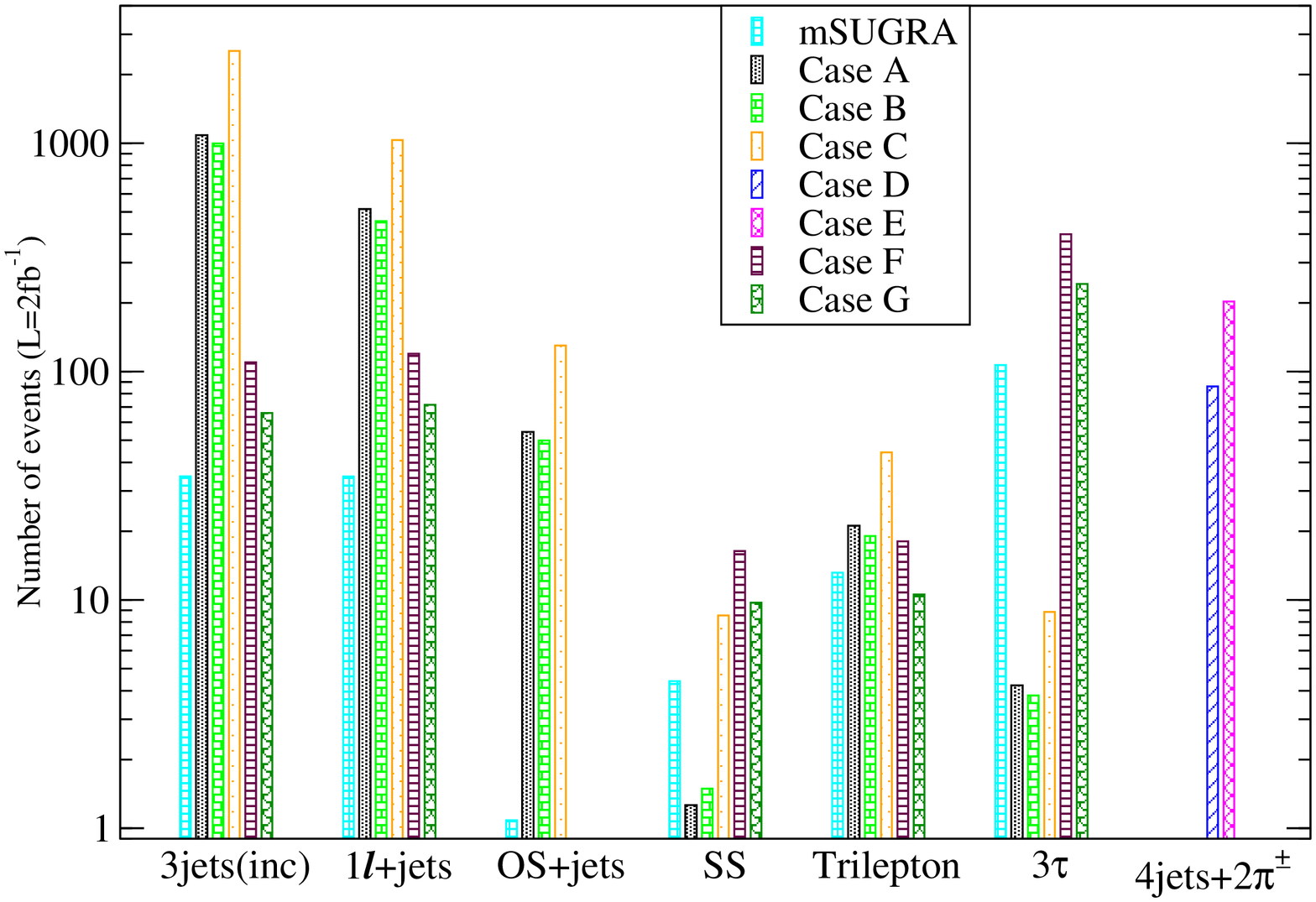}
\epsfxsize=15cm
\end{center}
{\caption{\label{fig:signature}\footnotesize {\bf Number of
superpartner events of different signatures for different models
at the Tevatron with $2\mbox{fb}^{-1}$.} These numbers are based
on counting topologies from {\tt Pythia} at the parton level with
no kinematic or geometric cuts. Every signature has missing
energy. From left to right, the signatures are: (1) inclusive
multi-jets $n_{jets}\ge 3$, (2) one lepton plus $n_{jets}\ge 2$,
(3) opposite sign dileptons plus $n_{jets}\ge 2$, (4) same-sign
dileptons, (5) trilepton, (6) 3 taus plus jets [before decaying
the taus], and (7) 4 jets plus 2 soft, isolated, high impact
parameter charged pions [some pions are like--signed].
For signatures (4)-(6), no requirement is made on the number of
jets.
A background analysis must of course be done to be sure any given
channel is detectable, but models with hundreds of events are
presumably detectable for the first two signatures, and models
with tens of events for the rest. The same-sign dilepton channel
has smaller backgrounds: even a handful of clean events may
constitute a signal. Here the model labeled ``mSUGRA'' is the
cMSSM Point~B of~\cite{bench}.}}
\end{figure}

\begin{figure}
\begin{center}
\includegraphics[scale=0.5,angle=0]{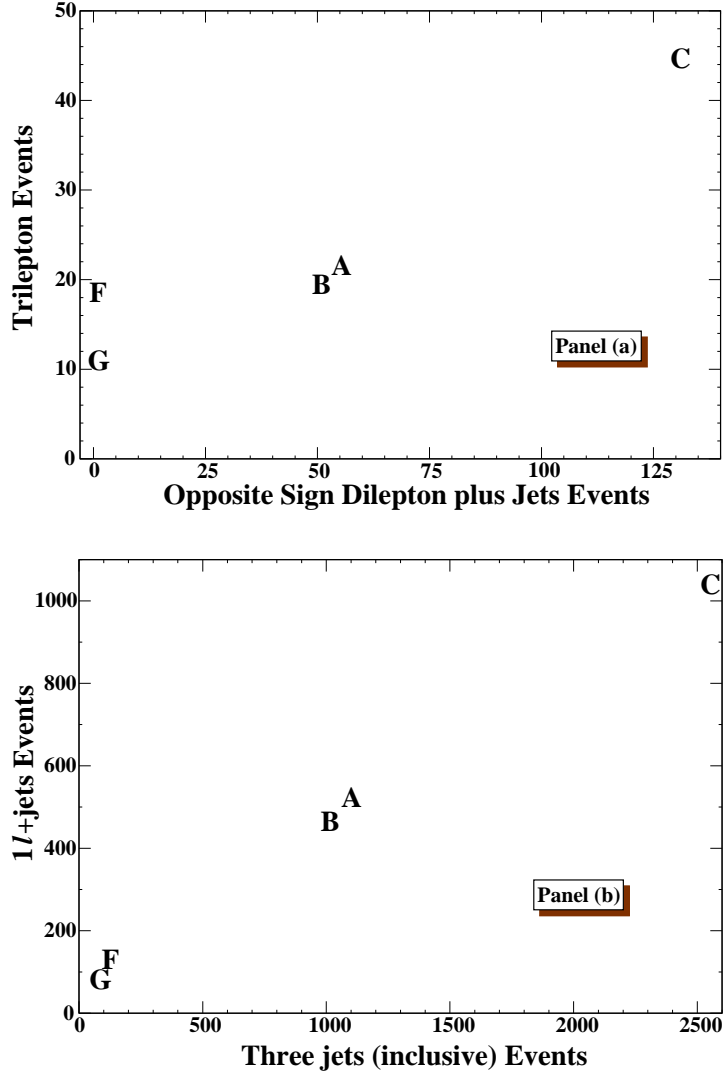}
\end{center}
{\caption{\footnotesize\label{fig:ratio}{\bf Correlations between
the number of events in different signatures for
$2\mbox{fb}^{-1}$}.  Cases~C,~FG, and AB are well separated from
each other. By comparing several such signatures, many models can
be identified at the Tevatron alone. A and B are difficult to
separate at the Tevatron but since their $\mu$, squark and slepton
masses are different, they should have different predictions for
low energy experiments ({\em e.g.} in rare decays or $g_\mu-2$)
and for LHC. Case G can be distinguished from others by the
3$\tau$ signature. Cases~D and~E (not shown here) have their
unique 4 jets plus 2 isolated soft, high impact parameter pions
signature with quite different event rates.}}
\end{figure}

Figure~\ref{fig:signature} shows naive estimates of numbers of
events in 2 fb$^{-1}$ integrated luminosity for various models and
various inclusive signatures.
The signature of these models are calculated using {\tt PYTHIA}
\cite{Sjostrand:2000wi}, but only at the generator level: no
geometric or kinematic cuts or triggering efficiencies are
applied, no jet clustering is performed, tau leptons are not
decayed, etc. The event numbers are only meant to illustrate the
generic features of each model and demonstrate the experimental
challenges.
In general, there will be too few events from any single exclusive
process to isolate it by cuts and observe a clean signal, but
significant excesses could be established in several inclusive
signatures. In all these cases there are of course backgrounds,
but typically the backgrounds are not so large as to prevent the
experiments from establishing an excess. Note that each model has
a different pattern, and it could be possible to learn quite a bit
about the underlying physics from the relative sizes of different
inclusive signals, as illustrated in Figure~\ref{fig:ratio}.
Despite the fact that the event numbers are based on
unsophisticated estimates, the resulting correlations are quite
distinct and should be robust under more detailed analyses.
Once a signal of beyond the Standard Model physics is established
it will be an exciting challenge to determine which superpartners
are being produced, their masses and branching ratios, and their
implications for the underlying theory.

Some specific features and signatures are worth noting. First, the
negative gluino mass of the two moduli-dominated models D and~E is
physical and observable~\cite{KaMrWa00}. Next for these models the
LSP is predominantly a wino that is almost degenerate with the
lightest chargino so that the dominant decay mode of the chargino
is $\tilde W^\pm\to\tilde W^0 \pi^\pm$~\cite{GhGiWe99}. This is
quite similar to the anomaly-mediated supersymmetry breaking
models, but in AMSB the ratio of gaugino masses $M_1:M_2:|M_3|$ is
approximately $2.8:1:7.1$ so that the gluino is very heavy and out
of reach for the Tevatron. Thus in the usual anomaly-mediated
cases the chargino pair can only be produced directly and not from
gluino decay. In our cases~D and~E, however, gluino masses are
$351 \GeV$ and $325 \GeV$, respectively, so the cross section for
gluino pair production is quite large. The gluino has the decay
mode $\tilde g \to\tilde W^\pm  q q^\prime$ with a branching ratio
about 50\%, followed by $\tilde W^\pm\to\tilde W^0 \pi^\pm$. Thus
there will be large missing transverse energy with four jets plus
two soft, high impact parameter pions. Since the chargino $\tilde
W^\pm$ emerges from gluino decay it is quite energetic so the pion
will also be reasonably energetic and give a good signature to
detect these models. Furthermore, gluino production will lead to
pairs of like--signed pions in these events.  The number of such
events expected in cases~D and~E are shown in the last set of
columns in Figure~\ref{fig:Tevspec}. These are interesting new
collider signatures for the Tevatron, and for AMSB at the LHC,
that to our knowledge have not been studied previously.


For models~F and~G the stau $\tilde \tau_1$ is lighter than
$\tilde N_2$ and $\tilde C_1$, so
$\widetilde{N_{2}}\rightarrow\tilde{\tau}\tau$ and
$\widetilde{C_{1} }\rightarrow\tilde{\tau}\nu_{\tau}$ dominates,
leading to a large three tau signal and reducing the trilepton
rate. Although model~C also has many trilepton events, the reason
is different. For model~C, the  $\tilde N_2\tilde C_1$ cross
section is quite large, but the leptonic branching ratios of
$\tilde N_2$ and $\tilde C_1$ are smaller. Model~C has jets plus
missing energy signatures while models~F and~G do not.

Detecting and studying these models presents interesting
challenges for experiments at the LHC, which, in principle, has
the kinematic reach to produce all superpartners. However, many of
the scalars are quite heavy, and will have small production rates.
The models with light gluinos have relatively little missing
transverse energy and large backgrounds. Models~F and~G have
heavier gluinos with mainly two--body decays dominated by
$\tilde{g}\to\tilde{b}\bar{b},\;\tilde{t}\bar{t}$. All of our
models have at least some superpartners that will be detected at a
$500\GeV$ linear collider. With a $520\GeV$ linear collider every
model allows the study of several superpartners, though clearly
not all.

Further discrimination between models can be obtained by studying
ratios of numbers of events of different types of signatures. An
example of this is displayed in Figure~\ref{fig:ratio} where two
different pairs of signatures from Figure~\ref{fig:signature} are
analyzed. By comparing several such signatures the underlying
physics of many models can be identified at the Tevatron alone.
While cases~A and~B are difficult to separate at the Tevatron, it
may be possible to distinguish between them by their different
predictions for low energy experiments or by their different
predictions for scalar masses accessible at the LHC. Case G can be
distinguished from the others by the 3$\tau$ signature. Case D and
E have their unique 4 jets plus 2 isolated soft, high impact
parameter pions signature with quite different event rates (86 and
201 respectively), so that they can be easily distinguished from
other cases and amongst themselves.

We have checked that all of our benchmark models are not
inconsistent with indirect constraints on superpartner masses. For
example, the SUSY contribution to the muon anomalous magnetic
moment for all models falls within the ``conservative'' bound
obtained in~\cite{MaWe02}, particularly if one favors the Standard
Model prediction based on tau decay data. If one prefers instead
the Standard Model prediction based on $e^+ e^-$ collider data
then all models can be made consistent with measurements at the
the $2\sigma$ level by simply increasing the value of $\tan\beta$.
As our focus has been on studying collider signatures for a range
of $\tan\beta$ values we have chosen not to tune our values in
this manner. The same is true for ${\rm Br}(b \to s \gamma)$ in
these models.

The thermal relic density of LSP neutralinos was computed for all
of these models using the {\tt DarkSUSY} program~\cite{Dark}. In
general, the $\Omega_{\rm LSP}{\rm h}^{2}$ results are very
sensitive to some parameters so we view any values below
$\Omega_{\rm LSP}{\rm h}^{2} \sim 2$ as satisfactory for the
purposes of a benchmark model. For example, in model~A the relic
density as computed by {\tt DarkSUSY} is $\Omega_{\rm LSP}{\rm
h}^{2} = 1.9$. However, lowering the top quark mass to $173.0
\GeV$ reduces this number to $\Omega_{\rm LSP}{\rm h}^{2} = 0.14$.
Alternatively, increasing the value of $\tan\beta$ from 10 to 12
changes this number to $\Omega_{\rm LSP}{\rm h}^{2} = 1.07$. For
both of the small modifications described above the superpartner
spectrum -- and thus, the collider signature of this model -- is
largely unchanged.  The sensitivity of the relic density in this
model (and model~B, the only other case where the LSP relic
density was larger than the cosmologically preferred region), is
due to the importance of chargino/neutralino coannihilations for
these models~\cite{BiNe01}.  Our focus in this work is not to
impose aggressive constraints on the MSSM parameter space but
rather to study the collider signatures of a representative sample
of theory-motivated models, so we have chosen not to adjust our
input parameters in such an artificial way. Consistency with all
indirect constraints on superpartners can be obtained by small
corrections to our benchmark points along any of the indicated
``model lines'' suggested in the Appendix.


\subsection{Fine Tuning}
\label{sec:tuning}

Any discussion of fine-tuning must necessarily involve certain
subjective statements. One commonly employed tool for comparing
models in a semi-quantitative way is the ``sensitivity parameter''
of Barbieri and Giudice~\cite{BaGi88} which measures the relative
change in the Z-mass when a high-scale input parameter is varied.
However, we believe that the degree of fine-tuning in a given
model may be more profitably thought of as divided between an
element that involves cancellations among various terms from the
soft supersymmetry breaking Lagrangian, and an element that
involves a measure of sensitivity arising from the overall scale
of supersymmetry breaking relative to the Z-mass scale. As was
pointed out some time ago~\cite{AnCa95} it is really the former
that is a measure of the fine-tuning in a given theory, while the
latter may often give misleading measures of tuning --
particularly when the {\em entire} parameter space of a particular
model implies a consistently high supersymmetry breaking scale as
characterized by the gravitino mass or scalar/gaugino masses. The
mere existence of a large scale in the theory need not necessarily
imply large fine-tuning, as was demonstrated for example
in~\cite{FeMaMo00a}, but the cancellation of large numbers against
one another to produce a much smaller number almost certainly does
if such cancellations can not be explained from the underlying
theory.

The degree of cancellation that a particular model of
supersymmetry breaking requires to obtain the correct Z-boson mass
can be expressed by simply expanding the formula that determines
$M_{Z}$ at the electroweak scale
\begin{equation}
\frac{M_{Z}^{2}}{2} = - \mu^{2}(t) + \( \frac{m^{2}_{H_D}(t) -
m^{2}_{H_U}(t) \tan^{2}\beta}{\tan^{2}\beta -1} \),
\end{equation}
with $t=\ln(\Lambda_{\UV}/Q)$, in terms of semi-analytic solutions
for the running parameters in terms of the input parameters at the
high scale $\Lambda_{\UV}$ and a given value of
$\tan\beta$~\cite{KaKi99}. For example, taking
$\Lambda_{\UV}=\Lambda_{\GUT}=2 \times 10^{16} \GeV$, the result
for $m_{\rm top}(M_Z) = 170$ and $\tan\beta=30$ is
\begin{eqnarray}
M_{Z}^{2}&=&-1.5 \mu^{2}(\UV) + 6.4 M_{3}^{2}(\UV) -0.4
M_{2}^{2}(\UV) + 0.0003 M_{1}^{2}(\UV) -1.2 m^{2}_{H_U}(\UV)
\nonumber \\ & & - 0.08 m^{2}_{H_D}(\UV) +0.8 m^{2}_{Q_3}(\UV) +
0.7 m^{2}_{U_3}(\UV) + 0.03 (m^{2}_{D_3}(\UV) + m^{2}_{L_3}(\UV) +
m^{2}_{E_3}(\UV)) \nonumber \\ & & +0.2 A_{t}^{2}(\UV)
 -0.6 A_{t}(\UV) M_{3}(\UV) -0.1
A_{t}(\UV)M_{2}(\UV) -0.002A_{t}(\UV)M_{1}(\UV) \nonumber \\ & &
+0.5M_{2}(\UV)M_{3}(\UV) +0.06M_{1}(\UV)M_{3}(\UV) +0.01
M_{1}(\UV)M_{2}(\UV) . \label{ztune2}
\end{eqnarray}

If the smallness of $M_Z = 91.187 \GeV$ is not to be the result of
a miraculous cancellation between large numbers then at least one
of the following must occur: either certain relations among the
soft terms and $\mu$ parameter must exist that guarantee
cancellations over a wide range of parameters, or each of the
individual terms in~(\ref{ztune2}) must be no more than a few
times $M_Z$ in size. The first could occur in theories of
supersymmetry breaking. Even the cMSSM predicts certain
``relations'' among soft terms that are postulated to hold over
all parameters: namely that gaugino masses and scalar masses are
unified. But in~\cite{KaLyNeWa02} it was argued that this alone is
not sufficient to prevent fine-tuning in the EWSB sector without
also postulating a robust relationship between $\mu$ and $M_3$
(the two most crucial parameters in determining the Z boson mass).
The conclusion drawn there was that the only reasonable way to
avoid unnatural cancellations in the determination of $M_Z$ is for
both $\mu$ and $M_3$ to individually be small. This implies that a
certain degree of nonuniversality in the gaugino masses is
beneficial in reducing EWSB fine-tuning while satisfying the
search limits from LEP. Note that the scalar masses
in~(\ref{ztune2}) are far less important in this regard.

For a given model we can determine everything in~(\ref{ztune2}),
apart from $\mu(\UV)$ itself, which has nothing to do {\em a
priori} with supersymmetry breaking, as a function of the value of
the coupling constant at the string scale $g_{\STR}^{2}$ (which we
can take to be $g_{\STR}^{2} \simeq 1/2$), the gravitino mass, and
a small number of free parameters related to the given model. Let
us take as an example the class of models from
Section~\ref{sec:theory2}. After choosing the initial scale and
the value of $\tan\beta$ the only free parameters are $m_{3/2}$,
$\delta_{\GS}$ and $\lang \re \; T \rang$.
Substituting~(\ref{modterms}) into~(\ref{ztune2}) with
$\delta_{\GS} = -9$ and factoring out the gravitino mass gives
\begin{equation}
\(\frac{M_{Z}}{100 \GeV}\frac{1\TeV}{m_{3/2}}\)^2 =
-1.5\(\frac{\mu(\UV)}{100 \GeV}\frac{1\TeV}{m_{3/2}}\)^{2} + 0.1
+0.7(t+\bar{t})\Eisen + 1.2(t+\bar{t})^2 G_{2}^{2}(t,\bar{t}).
\label{ztune9}
\end{equation}
The first constant is the contribution of the scalar masses
in~(\ref{ztune2}) with some addition from the anomaly-generated
loop corrections to the gaugino masses. The case for other values
of $\delta_{\GS}$ is quite similar. The fine-tuning arising from
cancellations in the soft supersymmetry breaking Lagrangian is
clearly controlled by the value of $\lang \re \; T \rang$ through
the combination $(t+\bar{t})\Eisen$. For values of $\lang \re \; T
\rang$ just larger than its self-dual point this combination is
negative and less than unity, thus providing a model that has a
very low degree of internal cancellation.

But of course this is only part of the story. The fact that all
parameters in~(\ref{ztune9}) are $\order(1)$ and roughly the same
size implies that no large cancellation is required in this model
to achieve the correct Z-mass {\em provided the gravitino mass
(scale) is approximately 1 TeV}. But since all soft supersymmetry
breaking terms are induced at the loop level, we expect $m_{3/2}
\sim 16\pi^2 M_z \sim 15 \TeV$. The need for such a large scale in
the theory is quite clear: the LEP limit on the Higgs mass of $m_h
\geq 114 \GeV$ implies, at least to a first approximation, that
some squark masses must be large in order to generate large
radiative corrections to the Higgs mass that are only
logarithmically sensitive to scale. This can be achieved when the
entire scalar sector of the theory is heavy at the high energy
scale, or alternatively the scalar masses can start small at the
high scale and the necessary large squark masses can be induced
through RG evolution by a large gluino mass.

Thus some degree of tuning arising from the overall scale will
likely be present in many low-energy models of supersymmetry
breaking, but those with large gluino masses also give rise to
troubling cancellations. Issues of tuning in the scale of soft
terms are intimately related to the question of generating the
supersymmetric $\mu$ parameter and are beyond the scope of this
paper. We have not found models with small tunings associated with
the overall scale, though we look at a significant class of string
effective theories. Here we have chosen instead to be guided
by~\cite{KaLyNeWa02} and have sought models that are capable of
generating a sufficiently massive Higgs boson without introducing
large internal cancellations within the soft supersymmetry
breaking Lagrangian itself. Each of our models involves robust
relations among the soft terms in the theory, in a manner dictated
by the fundamental theory as in~(\ref{ztune9}), which reduces the
cancellations required in the soft supersymmetry-breaking sector
relative to the typically-studied universal models. In this
limited sense, then, we find these models to be more ``natural''
than their universal counterparts, though some degree of tuning
remains.